\documentclass[aps,twocolumn,amsmath,amssymb,amsfonts,floatfix,superscriptaddress,showkeys,showpacs]{revtex4-1}
\usepackage{dcolumn}  
\usepackage{multirow}
\usepackage[T1]{fontenc}
\usepackage[latin1]{inputenc}
\usepackage{dsfont}
\usepackage{verbatim}
\usepackage{bm} 
\usepackage{hyperref} 
\usepackage{cleveref} 
\usepackage{floatrow}
\usepackage{natbib}
\usepackage{graphicx}
\usepackage{epstopdf}
\usepackage{ifpdf} 
\usepackage{epsfig}
\usepackage{color}
\usepackage[usenames,dvipsnames]{xcolor}
\usepackage[caption=false]{subfig}
\usepackage[export]{adjustbox}
\usepackage[colorinlistoftodos]{todonotes}

\newcommand{\br}{{\mathbf{r}}}

\newcommand{\bu}{{\mathbf{u}}}
\newcommand{\bJ}{{\mathbf{J}}}
\newcommand{\cP}{{\mathcal{P}}}
\newcommand{\cL}{\hat{\mathcal{L}}}
\newcommand{\cM}{\hat{\mathcal{M}}}
\newcommand{\cJ}{{\mathcal{J}}}
\newcommand{\cA}{{\mathcal{A}}}
\newcommand{\cI}{{\mathcal{I}}}
\newcommand{\Pe}{{\textrm{Pe\,}}}

\newcommand{\be}{{\boldsymbol{e}}}
\newcommand{\del}{\nabla}

\newcommand{\bcJ}{{\boldsymbol{\cJ}}}

\begin{document}

\title{Active fluids at circular boundaries: Swim pressure and anomalous droplet ripening}

\author{Tayeb \surname {Jamali}}
\email{tayeb@ipm.ir}
\affiliation{School of Physics, Institute for Research in Fundamental Sciences (IPM), Tehran 19395-5531, Iran}
\author{Ali \surname {Naji}}
\email{a.naji@ipm.ir}
\affiliation{School of Physics, Institute for Research in Fundamental Sciences (IPM), Tehran 19395-5531, Iran}

\date{\today}

\begin{abstract}
We investigate the swim pressure exerted by non-chiral and chiral active particles on convex or concave circular  boundaries.  Active particles are modeled as non-interacting and non-aligning self-propelled Brownian particles. The convex and concave circular boundaries are used as models representing a fixed inclusion immersed in an active bath and a cavity (or container) enclosing the active particles, respectively. We first present a detailed analysis of the role of convex versus concave boundary curvature and of the chirality of active particles on their spatial distribution, chirality-induced currents, and the swim pressure they exert on the bounding surfaces. The results will then be used to predict the mechanical equilibria of suspended fluid enclosures (generically referred to as `droplets') in a bulk with active particles being present either inside the bulk fluid or within the suspended droplets. We show that, while droplets containing active particles and suspended in a normal bulk behave in accordance with standard capillary paradigms, those containing a normal fluid exhibit anomalous behaviors when suspended in an active bulk. In the latter case, the excess swim pressure results in non-monotonic dependence of the inside droplet pressure on the droplet radius. As a result, we find a regime of anomalous capillarity for a single droplet, where the inside droplet pressure increases upon increasing the droplet size. In the case of two interconnected droplets, we show that mechanical equilibrium can occur also when they have different sizes. We  further identify a regime of anomalous ripening, where two unequal-sized droplets can reach a final state of equal sizes upon interconnection, in stark contrast with the standard Ostwald ripening phenomenon, implying shrinkage of the smaller droplet in favor of the larger one. 
\end{abstract}


\maketitle

\section{Introduction}

Active fluids, comprising self-propelled micro- and/or nano-particles in fluid media, have been studied extensively in recent years, due both to their fundamental relevance in non-equilibrium soft matter and statistical physics \cite{Lauga:RPP009,ramaswamyreview,golestanian_review,Romanczuk:EPJ2012,Marchetti:RMP2013,Yeomans:EPJ2014,gompper_review,bechinger_review,ZottlStark_review} and to their experimental applications \cite{Paxton2006_review,cargo,robotic,sperm-carrying,janusmain,Ebbens2010_Pursuit}. Active particles exhibit an ability to take up energy from their environment and convert it into self-propelled motion by utilizing internal mechanisms such as solvent-induced catalytic surface reactions in synthetic Janus micro- or nanoparticles \cite{janusmain,Howse2007_Self-Motile,Ebbens2010_Pursuit}, and flagellar or ciliary movements in biological swimmers  \cite{goldstein_review,Lauga:ANNREVF2016}.

Of particular interest has been the behavior of active particles near confining boundaries. Because of their persistent motion (exhibiting a finite run length before rotational diffusion or random tumbles redirect particle trajectory),  and also because of their hydrodynamic coupling to the bounding walls (where applicable), active particles exhibit prolonged detention times and, thus, markedly increased near-wall accumulation \cite{Elgeti2016_Microswimmers,Elgeti2013_WallAccumulation,rusconi,Hernandez-Ortiz1,wallattraction,li2011accumulation,wallattraction3,Underhill_2014,ardekani,Schaar2015_PRL,Mathijssen:2016c}. 
Particle activity can thus result in excessive mechanical pressure on the confining boundaries. Being known as `swim pressure', this quantity, and the question wether it constitutes a state function (enabling possible thermodynamic analogies for the description of active fluids based only on their bulk properties) have become one of the most intensely scrutinized topics over the last few years~ \cite{Lion2014_Osmosis,Takatori2014_SwimPressure,yang_aggregation_2014,Fily2014_Dynamics,Mallory2014_Anomalous,Mallory2014_PRE,Smallenburg2015_PRE,Winkler_SM2015,Fily2015_DynamicsDensity,yan_swim_2015,Takatori_PRE2015,Solon2015_Nature,Solon2015_PRL,Yan_JFM2015,Wysocki_2015,Nikola2016_PRL,Ginot2015_PRX,Speck_PRE2016,Marconi2016_Pressure,Junot_PRL2017,Lee_SM2017,Paoluzzi_SR2016,Fily_JPA2018}.  In this context, the effects of boundary curvature has specifically been addressed (see, e.g., Refs. \cite{Mallory2014_Anomalous,Mallory2014_PRE,Fily2014_Dynamics,Winkler_SM2015,Fily2015_DynamicsDensity,Wysocki_2015,Yan_JFM2015,Smallenburg2015_PRE,Nikola2016_PRL,Junot_PRL2017,Lee_SM2017,Paoluzzi_SR2016}). As compared with flat walls, active particles accumulate more (less) strongly near concave (convex) boundaries, exerting larger (smaller) pressure on them.

Fluid enclosures such as droplets provide an interesting example of the situation where active particles are exposed to curved boundaries (see, e.g., Refs. \cite{Cates_PNAS2010,Tjhung_PNAS2012,Tjhung_PNAS2016,Vladescu_PRL2014,Paoluzzi_SR2016,Hennes_PNAS2017,Jin_2018} and references therein). Confining active particles inside a droplet comes with interesting consequences. These include net motility displayed by deformable droplets containing a polar active material with contractile (as in active gels such as actomyosin) or extensile activity (as in motile, or non-motile and yet active, bacteria) \cite{Tjhung_PNAS2012,Tjhung_PNAS2016}. 
Active emulsions constitute another relevant area, where chemical activity due to reacting components -- rather than active self-propulsion -- generates intriguing droplet growth and division reminiscent of the cell cycle  \cite{Julicher:Nature2016,Julicher_PRE2015,Julicher_NJP2017,Golestanian_Nature2017,Cate_2018review}. These processes provide a mechanism  for stabilization of emulsions against coarsening or the so-called Ostwald ripening \cite{Israelachvili2015,deGennes_Capillarity_2004,Ostwald1900}, the process by which larger droplets grow (indefinitely) at the expense of the smaller ones. The latter is caused by interfacial capillary stress that produces higher inside pressure within the smaller droplets; hence, triggering content transfer from smaller to larger droplets (due, e.g.,  to solubility of the content material, leading to finite diffusive fluxes in the surrounding bulk, and/or to droplet coalescence). In the example of chemically active droplets with first-order reactions, not only is the Ostwald ripening suppressed, but the emulsion is also shown to become mono-disperse over time  \cite{Julicher:Nature2016,Julicher_PRE2015,Julicher_NJP2017}; both effects being of importance in technological applications \cite{Cate_2018review} and in the formation of liquid-like subcellular domains such as centrosomes \cite{Hyman_2014}. 
 
To our knowledge, despite several well-established routes available for stabilizing droplets in non-active systems \cite{Cate_2018review}, potential strategies to suppress Ostwald ripening in the presence of self-propelled particles remain largely unexplored. In fact, in one-component suspensions of (non-aligning) self-propelled particles, finite domains (formed as a result of the interplay between motility-induced aggregation and intrinsic active fluctuations in the system) are found to coarsen to a state of complete phase separation \cite{Marenduzzo_MIPS2015,Cates_Nature2014,Tailleur2008_StatisticalMechanics,Fily2012_PhaseSeparation,Redner2013_Structure,Buttinoni2013_DynamicalClustering,Cates2015_Motility-Induced,Liu2016}. The same applies to binary mixtures of active and non-active particles, where it is shown that even suitably prepared circular droplets of non-active particles surrounded by active self-propellers are unstable to macroscopic phase separation \cite{Stenhammar_PRL2015,Wittkowski_NJP2017}. Such droplets can however be stabilized individually by imposing interfacial stress (e.g., by means of an enclosing, or encapsulating, semipermeable surface). It remains unclear whether stabilized droplets in an active bulk will still follow the Ostwald-ripening scenario, when they come in contact and can undergo coalescence. Our aim is to address this question by elucidating the underlying physics, which can be established most clearly by investigating the inside pressure of a single droplet as well as the fate of two interconnected droplets on a mean-field level. We shall consider two conversely designed cases of (i) non-active droplets with normal fluid content suspended in a bulk containing self-propelled particles, and (ii) active droplets containing self-propelled particles suspended in a normal fluid. 

While the swim pressure can significantly increase the inside droplet pressure in both of the mentioned cases, drastic qualitative changes transpire in case (i), where the interplay between the interfacial tension and the swim pressure exerted on the external surface of the droplets leads to a non-monotonic  inside-pressure profile as a function of the droplet radius. This is at odds with the standard capillary paradigm, giving monotonically decreasing inside pressure with the droplet size. Hence, we predict a regime of {\em anomalous droplet ripening} over a wide range of realistic parameter values, where initially unequal-sized droplets are driven toward a stable state of mechanical equilibrium with equalized sizes. 

We also provide an inclusive analysis of the behavior of the swim pressure on convex and concave circular boundaries, corroborating some of the previous results within minimal models of non-chiral  active particles \cite{Mallory2014_Anomalous,Mallory2014_PRE,Fily2014_Dynamics,Winkler_SM2015,Fily2015_DynamicsDensity,Wysocki_2015,Yan_JFM2015,Smallenburg2015_PRE}. In particular, we give a detailed account of particle chirality effects, which have received mounting interest over the last several years \cite{Golestanian:PRE2010,takagi,takagi2013,Bechinger:PRL2013,Xue:EPJST2014,Xue:EPL2015,Wykes_2016,Crespi:2013,Teeffelen:PRE2008,Reichhardt:2013,Mijalkov:2015,Lowen:EPJST2016,Popescu_PRL2016,Liebchen_2016,Simmchen_Nature2016}, even though they have not been analyzed within the present context so far. We discuss the direct relation between the chirality-induced current around the circular boundaries and the chirality-induced reduction in swim pressure exerted on them. It is shown systematically that the properties (including the swim pressure) of a system of chiral self-propelled particles consistently converge to those of a corresponding non-active (or, equilibrium) system, when the chirality strength is taken to infinity.

\begin{figure}[t!]
\centering
\includegraphics[width=0.95\linewidth]{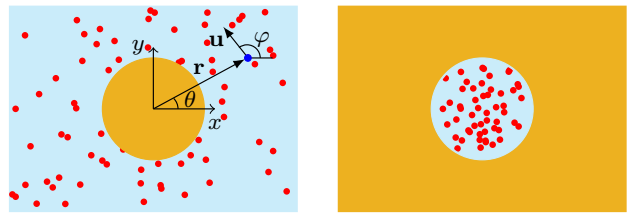}
\caption{Schematic view of active particles  (small red disks) at convex and concave boundaries, representing a disk-shaped inclusion (left) and a circular cavity (right), respectively. These geometries are also used to address stabilized droplets in a bulk fluid with active particles being present either inside or outside the droplet. }
\label{fig:schematic}
\end{figure}

\section{Model}
\label{sec:model}

To disentangle the effects of the swim pressure on curved boundaries from other possible effects such as collective alignment \cite{ramaswamyreview,Marchetti:RMP2013,gompper_review}, hydrodynamic interactions \cite{Hernandez-Ortiz1,wallattraction,li2011accumulation,wallattraction3,Underhill_2014,ardekani,Schaar2015_PRL,Mathijssen:2016c}, and steric particle layering  \cite{Mallory2014_Anomalous,Ni2015_Tunable,Mahdi2017_ChiralSwimmers}, we focus on the so-called minimal model of ideal (non-interacting) active Brownian particles in their commonly used two-dimensional formulation \cite{Romanczuk:EPJ2012}. The active particles are considered in contact with a circular interface of radius $R$, which represents the bounding surface of a fixed, disk-shaped {\em inclusion} immersed in the active bath, or that of a {\em cavity} enclosing the active particles  (see Fig. \ref{fig:schematic}).  While the interface can generally be taken to be permeating or non-permeating to the background solvent, it is assumed to be non-permeating to active particles. The particles are repelled from the interface by the harmonic potential 
\begin{equation}
V(\br)=\frac{1}{2}k(|\br|-R)^2 \mathcal{B}(\br),
\end{equation}
 with $k>0$ being a sufficiently large constant to represent a nearly hard boundary, and $\mathcal{B}(\br)$ the blip function 
\begin{align}
\label{eq:blip}
\mathcal{B}(\br) =  \left\{
\begin{array}{ll}
     \Theta(R-|\br|)&\qquad   {\textrm{(inclusion)}},\\
     \\
     \Theta(|\br|-R)&\qquad   {\textrm{(cavity)}}, 
\end{array}
\right.
\end{align}
where $\Theta(\cdot)$ is the Heaviside step function. 

The stochastic motion of an individual active particle is described by the customarily used, two-dimensional Langevin-type equations \cite{Romanczuk:EPJ2012,bechinger_review} 
\begin{align}
\label{eq:Langevin}
\dot{\br} &=  v\bu-\mu_t \del V + \sqrt{2D_t}\, \boldsymbol{\eta}(t), \\
\label{eq:Langevin_rot}
\dot{\varphi} &= \omega + \sqrt{2D_r}\, \zeta(t),
\end{align}
where ${\mathbf r}(t) = (x(t), y(t))$ is the instantaneous position vector of the active particle measured from the geometric origin, and $\bu(t) = (\cos\varphi(t),\sin\varphi(t))$ its instantaneous orientation unit vector, pointing in the direction of  self-propulsion; the latter is parameterized by the orientational (azimuthal) angle $\varphi(t)$ with respect to the $x$-axis  (Fig. \ref{fig:schematic}). The active particles self-propel at a constant linear speed $v$ and are generally chiral, possessing a constant intrinsic angular velocity of $\omega$ that can originate from unbalanced self-propelling torques. The particle dynamics is also subject to independent translational and rotational, Gaussian, white noises $\boldsymbol{\eta}(t)$ and $\zeta(t)$, with zero mean and unit variance, where $D_t$ and $D_r$ denote the corresponding translational and rotational diffusivities, respectively, and $\mu_t$, the translational mobility. 

The choice of white noise terms and the Einstein-Smoluchowski-Sutherland relation,
$D_t = \mu_t k_{\mathrm{B}}T$, where $k_{\mathrm{B}}$ is the Boltzmann constant and $T$ the ambient temperature, ensure that the properties of the system reduce to their corresponding equilibrium, Boltzmann weighted, states when the active self-propulsion is switched off. 
On the other hand, our assumption of non-interacting particles (as applicable to sufficiently dilute suspensions; see, e.g., Refs. \cite{Solon2015_Nature,Nikola2016_PRL,saintillan2015bookchapter} and references herein) allows for a convenient continuum probabilistic approach based on a  local Smoluchowski-Fokker-Planck equation. 

Equations \eqref{eq:Langevin} and \eqref{eq:Langevin_rot} can standardly be mapped \cite{Risken_FP_1996} to the Smoluchowski-Fokker-Planck equation for the joint probability density function (PDF) of active particles, $\cP(\br,\varphi; t)$, over the position-orientation space defined by the coordinates $\br=(x,y)$ and $\varphi$ as
\begin{align} 
\label{eq:master_eq}
\nonumber
\partial_t\cP &=  D_t\del^2\cP + D_r\partial_\varphi^2 \cP  \\
  & -\del\cdot\left[(v\bu  - \mu_t \del V)\cP\right] - \omega\partial_\varphi \cP,
\end{align}
where  $\del \equiv \partial/\partial \br$ and $\partial_\varphi \equiv  \partial/\partial \varphi$. 

The steady-state swim pressure exerted on the circular interfaces can be obtained from the number density profile of active particles, which itself follows from the steady-state PDF, $\cP_0(\br,\varphi)$; the latter can be determined numerically from the solution of Eq.~(\ref{eq:master_eq}) for $\partial_t\cP=0$. The resulting equation can be expressed in the coordinate space spanned by  $(\br,\varphi) = (r,\theta,\varphi)$. Rotational symmetry, however, dictates that $\cP_0 = \cP_0(r,\theta,\varphi)$ must depend on its angular arguments only through the combination $\psi\equiv \varphi-\theta$; that is,  $\cP_0=\cP_0(r,\psi)$. Therefore, since $\partial_\varphi  \cP_0 = -\partial_\theta  \cP_0= \partial_\psi  \cP_0$, we can conveniently express the steady-state form of Eq. (\ref{eq:master_eq}) as 
\begin{equation}
\label{eq:master_eq_polar}
\del_\psi\cdot\bcJ =  \frac{1}{ r}  \partial_{ r}( r {\cJ}_{ r}) +  \frac{1}{ r} \partial_\psi{\cJ}_\psi=0, 
\end{equation}
where the subscript $\psi$ is used to indicate that the divergence is taken in the polar coordinates $(r,\psi)$. Also, the probability current density $\bcJ=(\cJ_r,\cJ_\psi)$ reads 
\begin{align}
\label{eq:current density components}
\cJ_r &= (v\cos\psi-\mu_t \partial_r V)\cP_0-D_t\partial_r \cP_0, \\
\cJ_\psi &= (-v\sin\psi+r \omega) \cP_0 - \frac{1}{r}(D_t+D_r r^2) \partial_\psi \cP_0.
\end{align}

Before proceeding with our numerical results, we follow the methods developed in Refs.  \cite{Solon2015_Nature,Solon2015_PRL,Nikola2016_PRL} to derive semi-analytical relations facilitating further insight into the different factors that contribute to the swim pressure in the considered geometries. Since the above two cases turn out to be similar in our formulation, we first focus on the case of a disk-shaped inclusion and then generalize the final results directly to the case of a circular cavity.

\section{Swim pressure and current density}
\label{sec:swimm_pressure}

\subsection{Inclusion}
\label{subsec:inclusion_results_formulation}

The force exerted on the inclusion upon contact by an individual active particle is given by $\partial_r V(r)$. In the continuum description, the force per unit area at radial distance $r$ from the origin is given by $\rho\,\partial_r V$, where $\rho(r)$ is the local, active-particle number density 
\begin{equation}
\label{eq:number_density}
\rho(r) = \int_0^{2\pi} \cP_0(r,\psi)\,{\mathrm{d}}\psi. 
\end{equation}
The total force exerted on a shell of radius $r$ and thickness ${\mathrm{d}}r$ is thus given by $\rho\,(2\pi r {\mathrm{d}}r) \partial_r V$ and, hence, the pressure
\begin{equation}
\label{eq:pressure_inclusion}
P = \int_{\Lambda}^0 \rho \partial_r V\,{\mathrm{d}}r. 
\end{equation}
The radial integration is taken from a conventionally chosen point, $r=\Lambda$, in the {\em bulk}, where the density can be taken as constant $\rho_0=\rho(\Lambda)$, to the center of the inclusion at the origin, $r=0$, where the density is zero, $\rho(0)=0$ (these assertions will be verified numerically  in Section \ref{subsec:inclusion_results}). Even though $\partial_r V$ vanishes outside the inclusion $r>R$ (see Eq. \eqref{eq:blip}), taking the upper bound of the integral in the bulk (rather than on the inclusion surface) enables us to relate the pressure, $P$, to the predefined bulk density, $\rho_0$. This can be made clearer by writing Eq. (\ref{eq:pressure_inclusion}) in the form
\begin{equation}
\label{eq:pressure_1}
P = \left(\frac{D_t}{\mu_t}\right) \rho_0 + \frac{v}{\mu_t}\int_\Lambda^0  m_1(r) \,{\mathrm{d}}r.
\end{equation}
The above equation is obtained by making use of the first (angular) integral of  Eq.~(\ref{eq:master_eq_polar}) and by defining the harmonic moments of the PDF as
\begin{equation}
\label{eq:harmonic moments}
m_l(r) = \int_0^{2\pi} \cos(l\psi) \cP_0(r,\psi)\,{\mathrm{d}}\psi, 
\end{equation}
in which case, $l=0$ gives the steady-state number density of active particles, $m_0(r) = \rho(r)$. Equation \eqref{eq:pressure_1} comes with the  numerical advantage that it avoids computing the derivative of the potential that can be rapidly varying across the interfacial region. 

In the special case of non-interacting active particles next to a \emph{flat} wall, the pressure can be calculated in terms of the bulk density directly and without having to calculate the steady-state PDF $\cP_0(r,\psi)$; this is because the integrand in Eq.~(\ref{eq:pressure_inclusion}) or, equivalently, $m_1$ in Eq. \eqref{eq:pressure_1}, can be written as a  complete derivative \cite{Solon2015_Nature,footnote1}. This property does not hold in the case of a circular boundary. We can nevertheless find a useful expression for the swim pressure by expressing it in terms of the following three contributions (see Appendix~\ref{sec:app_Pressure_Equation} for details)
\begin{equation}
\label{eq:P_decomp}
P \simeq P_f + P_{ch} + P_{ex}, 
\end{equation}
where
\begin{align}
\label{eq:pressure_inclusion_2a}
P_f = \left(\frac{v^2}{2\mu_t D_r} + \frac{D_t}{\mu_t} \right)\rho_0
\end{align}
 is the standard swim pressure of non-chiral active particles on a flat wall \cite{Solon2015_Nature}, and we have 
\begin{align}
\label{eq:pressure_inclusion_2b}
&P_{ch} =-\frac{v\omega}{\mu_t D_r} \int_0^{2\pi} \!\!\!\int_\Lambda^0  \sin\psi\,\cP_0(r,\psi)\,{\mathrm{d}}r\,{\mathrm{d}}\psi, \\
\label{eq:pressure_inclusion_2c}
&P_{ex}= \frac{v}{\mu_t D_r} \int_\Lambda^0 \big(\!-vm_2+\mu_t m_1 \partial_r V+D_t \partial_r m_1 \big)\frac{ {\mathrm{d}}r}{r}. 
\end{align}
Here, $P_{ch}$ is a term with an explicit dependence on the active-particle chirality, and $P_{ex}$ is an excess contribution associated with the interfacial curvature and vanishes when the latter is set to zero ($R\rightarrow \infty$). It is important to note that, although $P_{ch}$ and $P_{ex}$ can be present independently of one another, the effects due to particle chirality and interfacial curvature are, in general, intertwined and non-additive, and enter implicitly in both $P_{ch}$  and $P_{ex}$ through the PDF $\cP_0(r,\psi)$. 

It turns out that $P_{ch}$ is always {\em negative} and, as such, reduces the swim pressure exerted on the boundary as compared with that of a non-chiral system. This is because chiral active particles can produce {\em net currents} rotating around circular boundaries. The steady-state particle current density $\bJ = J_r \be_r + J_\theta \be_\theta$ can be obtained by integrating Eq. (\ref{eq:master_eq_polar}) over the orientational degree of freedom (see Appendix \ref{sec:app_Current_Density} for details), giving $J_r=0$ and 
\begin{equation}
\label{eq:radial_current_density1}
J_\theta(r) = v \int_0^{2\pi} \sin\psi\, \cP_0(r,\psi)\,{\mathrm{d}}\psi. 
\end{equation} 
It is thus evident that the chirality-induced reduction in the swim pressure, $P_{ch}$, is directly related to the rotational current density, $J_\theta(r)$, as
\begin{align}
\label{eq:pressure_current_inclusion}
P_{ch} =-\frac{\omega}{\mu_t D_r} \int_\Lambda^0 J_\theta(r)\,{\mathrm{d}}r \equiv -\frac{\omega}{\mu_t D_r} I, 
\end{align}
where $I$ is  the total current of chiral active particles around the inclusion. As we show in Section \ref{sec:results}, the total current has the same sign as $\omega$, ensuring that $P_{ch}<0$. 

\subsection{Cavity}
\label{subsec:cavity_results_formulation}

When active particles are confined in a circular cavity, one can follow steps similar to those discussed in the case of an inclusion. In the case of a cavity, one should choose the appropriate form of the blip function from Eq. (\ref{eq:blip}) and take the radial integrals from the geometric center of the cavity, representing the `bulk', where the particle density is expected to be constant, $\rho_0=\rho(0)$ (as will be confirmed numerically later), to a point deep inside the wall, where the particle density is zero. Thus, the swim pressure on the cavity is obtained as
\begin{equation}
\label{eq:cavity_pressure_1}
P = \int_0^{\infty} \rho \partial_r V\,{\mathrm{d}}r.  
\end{equation}
The pressure can be written again as $P\simeq P_f + P_{ch} + P_{ex}$, with the different terms given by the same expressions as in Eqs. \eqref{eq:pressure_inclusion_2a}-\eqref{eq:pressure_inclusion_2c} with the only difference being that the integrals over $r$ need to be taken over the interval  $[0, \infty)$, similar to the one expressed in  Eq.~(\ref{eq:cavity_pressure_1}). The same applies to the integral expression in Eq. \eqref{eq:pressure_current_inclusion}. The term $P_{ch}$ also turns out to be negative, indicating a reduction in the swim pressure exerted on the cavity, when particle chirality is switched on.

\floatsetup[figure]{style=plain,subcapbesideposition=top}
\begin{figure*}[t!]
\sidesubfloat[]{%
  \label{sfig:ObjectPDF_Pe=10}
  \includegraphics[width=0.29\linewidth]{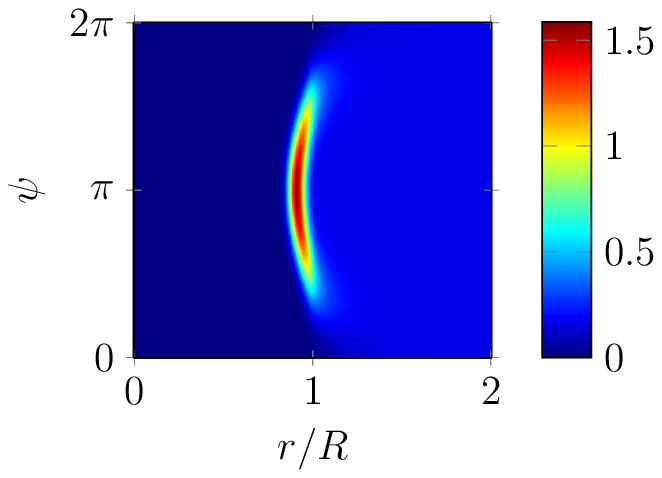}%
}\hfill
\sidesubfloat[]{%
  \label{sfig:ObjectDensity_Pe=10}
  \includegraphics[width=0.28\linewidth]{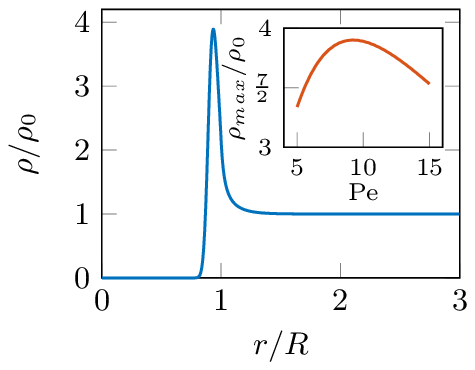}%
}\hfill
\sidesubfloat[]{%
  \label{sfig:ObjectPressure_PeDependency}
  \includegraphics[width=0.29\linewidth]{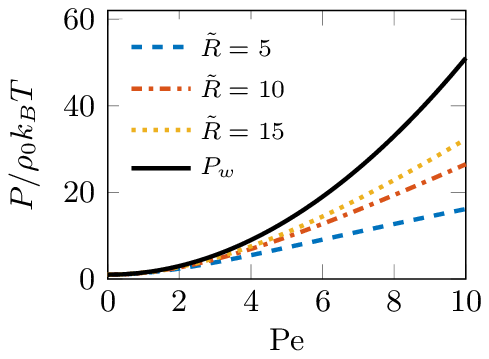}%
}\vspace{4mm}
\sidesubfloat[]{%
  \label{sfig:ObjectPressure_RDependency}
  \includegraphics[width=0.29\linewidth]{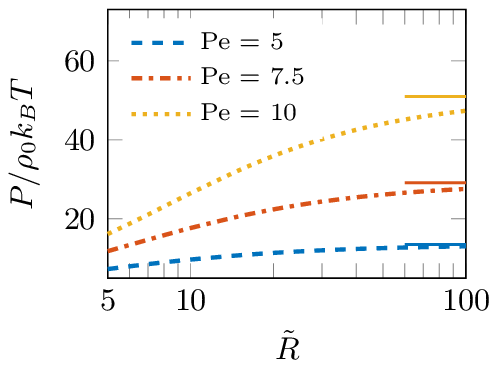}%
}\hfill
\sidesubfloat[]{%
  \label{sfig:ObjectPressure_PeDependency_Chiral}
  \includegraphics[width=0.29\linewidth]{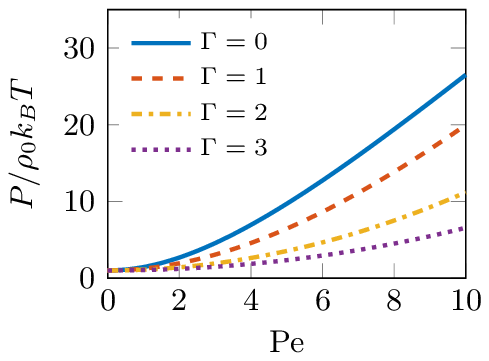}%
}\hfill
\sidesubfloat[]{%
  \label{sfig:ObjectPressure_ChiralityDependency}
  \includegraphics[width=0.29\linewidth]{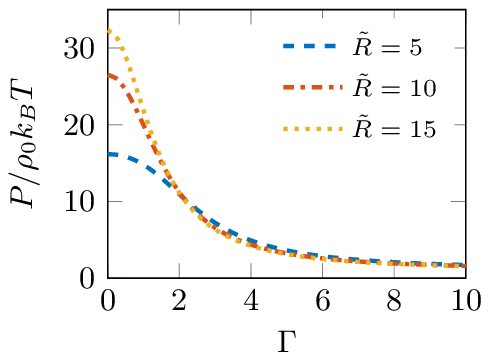}%
}
\caption{(a) Rescaled steady-state PDF, $\cP_0(r,\psi)/\rho_0$, of non-chiral ($\Gamma=0$) active particles around a nearly hard, disk-shaped inclusion immersed in the active bath is shown in the coordinate plane $(r/ R, \psi)$ for $\Pe = 10$ and $\tilde R = 10$. (b) The corresponding density profile of non-chiral active particles $\rho(r)$, divided by the bulk density $\rho_0$, as a function of the radial distance, $r/ R$, from the center of the inclusion. The inset shows the non-monotonic behavior of $\rho_{max}$ as a function of the P\'eclet number, $\Pe$. (c) Swim pressure of non-chiral active particles exerted on the inclusion surface as a function of $\Pe$ for three different rescaled radii $\tilde R=5,10$, and $15$, compared with the corresponding value obtained for a flat wall (black solid curve). (d) Swim pressure of non-chiral active particles as a function of the rescaled inclusion radius for three different values of $\Pe$ as indicated on the graph. (e) Swim pressure of chiral active particles for fixed $ \tilde R = 10$ and different values of the chirality strength parameter, $\Gamma$, as shown on the graph. (f) Swim pressure as a function of $\Gamma$ for fixed  $\Pe=10$ and three rescaled radii as shown on the plot. 
}
\label{fig:object_case}
\end{figure*}

\subsection{Non-dimensionlization}
\label{subsec:parameters}

In order to solve Eq.~(\ref{eq:master_eq_polar}) for the steady-state PDF using numerical methods, this equation is cast in a dimensionless form as discussed in Appendix \ref{sec:app_num_scheme}. We choose the characteristic length-scale of the system as $a=(D_t/D_r)^{1/2}$. It follows that the problem can be described by the dimensionless parameters, 
\begin{equation}
 \Pe=\frac{v}{aD_r},\quad \Gamma=\frac{\omega}{D_r},\quad \tilde R=\frac{R}{a}, \quad \tilde k=\frac{ka^2}{k_{\mathrm{B}} T}, 
\end{equation}
being the (swim) {\em  P\'eclet number} and the {\em chirality strength parameter} of active particles, the rescaled radius of curvature of the inclusion/cavity, and the rescaled interfacial potential strength,  respectively.  We fix  $\tilde k=10$, and   study the representative behavior of the system by varying the mentioned parameters over the ranges $\Pe=0 - 10$,  $\Gamma=0 - 10$, and $\tilde R= 5-100$. These values fall within the experimentally accessible ranges of parameters; for instance,  using typical Stokes diffusivities with $D_t \simeq 0.22\, \mu{\mathrm{m}}^2\cdot{\mathrm{s}}^{-1}$ and $D_r \simeq 0.16\,{\mathrm{s}}^{-1}$  in an aqueous medium, we find $a\simeq 1.2\, \mu{\mathrm{m}}$,  $v\simeq 0-2\, \mu{\mathrm{m}}\cdot{\mathrm{s}}^{-1}$ and $R\simeq 6-120\,  \mu{\mathrm{m}}$ \cite{bechinger_review}. Active particle exhibit a wide range of smaller and larger rotational diffusivities and self-propulsion speeds \cite{bechinger_review} (e.g., $D_r \simeq 1\,{\mathrm{s}}^{-1}$, $a\simeq 2\, \mu{\mathrm{m}}$ and  $v\simeq 20\, \mu{\mathrm{m}}\cdot{\mathrm{s}}^{-1}$ \cite{takagi}, giving $\Pe\simeq 10$), and also a wide range of values for the chirality strength. Examples of the latter include curved self-propelled rods (with a distribution of values for $\Gamma$, including  $\Gamma\simeq 1-5$) \cite{takagi,takagi2013}, Janus doublets ($\Gamma\simeq 8-26$) \cite{Golestanian:PRE2010}, active L-shaped particles ($\Gamma\simeq 225-435$) \cite{Bechinger:PRL2013}, and self-assembled rotors of two and three tripartite metallic rods ($\Gamma\simeq 2-13$) \cite{Wykes_2016}. 

\section{Results}
\label{sec:results}

We start by considering the PDF, particle density and swim pressure first in the case of an inclusion and then in the case of a cavity (Sections \ref{subsec:inclusion_results} and \ref{subsec:cavity_results}). The chirality-induced current density will be discussed later in Section \ref{subsec:current}. 

\subsection{Inclusion}
\label{subsec:inclusion_results}

Let us first consider an inclusion immersed a bath of {\em non-chiral} active particles ($\Gamma=0$). 
Figure~\ref{fig:object_case}\subref{sfig:ObjectPDF_Pe=10} shows the typical form of the numerically obtained, rescaled steady-state PDF of active particles in dimensionless form, $ \cP_0( r, \psi)/\rho_0$, over the $(r/ R, \psi)$-plane. Here, we have  $\Pe  = 10$ and $\tilde R=10$. The figure shows a peak with high active particle probabilities at and around the coordinates $\psi=\pi$ and $r= R$. This indicates  accumulation of active particles near the inclusion boundary, consistent with the wall-accumulation behavior generally known to occur for active particles (see, e.g., Refs. \cite{Elgeti2016_Microswimmers,Elgeti2013_WallAccumulation}). The results in Fig.~\ref{fig:object_case}\subref{sfig:ObjectPDF_Pe=10} also indicate that the majority of active particles around the inclusion orient in the {\em normal-to-surface} (or radial) direction, pointing {\em toward} the inclusion center, characterized by $\psi=\pi$. Even though active particles exhibit no discernible angular preference away from the inclusion (light-blue region for large $r/R$), the loci of maximum-density regions (in red) display a non-trivial angular distribution at the proximity of the inclusion (with a broad angular spread in the range $\pi/2\lesssim \psi \lesssim 3\pi/2$ for the parameters in the figure). Hence, as expected, the rescaled radial density profile (see Eq. (\ref{eq:number_density}) and Fig.~\ref{fig:object_case}\subref{sfig:ObjectDensity_Pe=10}) vanishes for the most part within the inclusion, exhibits a pronounced peak at the inclusion surface (with a peak location slightly displaced inside the inclusion due to the nearly hard nature of the  repulsive interfacial potential), and then rapidly falls off to a constant bulk value going away from the inclusion surface (corroborating the assumptions made in Section \ref{subsec:inclusion_results_formulation}). The maximum density of active particles shows a {\em non-monotonic} dependence on the P\'eclet number, $\Pe$ (see inset in Fig.~\ref{fig:object_case}\subref{sfig:ObjectDensity_Pe=10}): It increases up to a certain value of $\Pe$ (which, for the given parameters with $\tilde R=10$, occurs at $\Pe\simeq 9.2$) and then decreases. Intuitively, an increased P\'eclet number increases the particle-inclusion collision probability but, at the same time, reduces the probability of residing on the surface (and, hence, of surface detention time \cite{Schaar2015_PRL}) in a given time interval. These competing factors underly the observed non-monotonic behavior, indicating a crossover from the {\em active Brownian regime} (where the particle run length, $\ell_{\textrm{run}}=v/D_r$, is larger than the non-active Brownian scale,  $a$, but smaller than the inclusion radius, i.e., $a<\ell_{\textrm{run}}\lesssim R$, or $1<\Pe\lesssim\tilde R$ in rescaled units) to the {\em ballistic regime} at large $\Pe$ (where $\ell_{\textrm{run}} > R$, or $\Pe>\tilde R$ in rescaled units).

The rescaled swim pressure acting on the inclusion follows from the properly rescaled form of Eq. \eqref{eq:pressure_1} and calculating the zeroth and the first harmonic moments of the PDF. The results are shown in Fig.~\ref{fig:object_case}\subref{sfig:ObjectPressure_PeDependency} as a function of the P\'eclet number for three different values of the rescaled inclusion radius. For comparison, we also show the swim pressure exerted on a flat wall, $P_f$, Eq. \eqref{eq:pressure_inclusion_2a} (black solid curve). As seen, the pressure is smaller for smaller radii of curvature and, for instance, for  $\Pe=10$ and $\tilde R=5$, we find a drop of more than 65\% in the pressure from its reference value at a flat wall. Since we are considering the regime of relatively large inclusion radii, the approximate decomposition of the swim pressure into the three contributing terms, Eq. \eqref{eq:P_decomp}, turns out to be very accurate (Appendix~\ref{sec:app_Pressure_Equation}). Hence, since the particles are non-chiral ($P_{ch}=0$), the reported deviation from $P_f$ in Fig.~\ref{fig:object_case}\subref{sfig:ObjectPressure_PeDependency} comes only from the contribution $P_{ex}$, which is {\em negative}. Although a decrease in the swim pressure on a convex surface can intuitively be expected due to the shorter surface detention times of active particles, our results show that the influence of surface curvature on the magnitude of the swim pressure can be quite significant. Furthermore, the larger the P\'eclet number, the stronger will be the variations of the swim pressure with the radius of curvature (see Fig.~\ref{fig:object_case}\subref{sfig:ObjectPressure_RDependency}),  before the pressure saturates to its limiting flat-wall value (shown in the figure by horizontal line segments of the same color).  

In the case of {\em chiral} active particles, the effects due to self-propulsion \cite{Xue:EPJST2014,Xue:EPL2015} and, in particular, the surface accumulation of active particles are expected to diminish \cite{Mahdi2017_ChiralSwimmers}. The behavior of the system in the limit $\Gamma\rightarrow \infty$ consistently converges to that of a corresponding non-active (or, in the present model, equilibrium) system as we systematically prove in  Appendix~\ref{sec:app_effective_Smoluchowski_Eq}. As a result, the swim pressure is expected to drop in the case of chiral particles and tend to its non-active limiting value, $P/(\rho_0k_{\mathrm{B}}T)=1$,  as the particle chirality, $\Gamma$, is increased to infinity. Nevertheless, for a wide range of finite and experimentally accessible values of $\Gamma$, such as those shown in Fig.~\ref{fig:object_case}\subref{sfig:ObjectPressure_PeDependency_Chiral}, one deals with an intermediate situation, where the swim pressure is significantly reduced. The reduction is around 75\% at $\Pe=10$, when $\Gamma$ is increased from 0 to 3 in Fig.~\ref{fig:object_case}\subref{sfig:ObjectPressure_PeDependency_Chiral}; in this case, the pressure drop is mainly due to $P_{ch}$ (the changes in $P_{ex}$ due to the variation in $\Gamma$ is relatively small).   Figure~\ref{fig:object_case}\subref{sfig:ObjectPressure_ChiralityDependency} demonstrates the convergence of the swim pressure data at different sets of parameters to a single curve as $\Gamma$ is increased, and then gradually to the equilibrium bulk pressure, confirming the aforementioned limiting behavior.

\floatsetup[figure]{style=plain,subcapbesideposition=top}
\begin{figure*}[t!]
\centering
\sidesubfloat[]{%
  \label{sfig:CavityPDF_Pe=10}
  \includegraphics[width=0.3\linewidth]{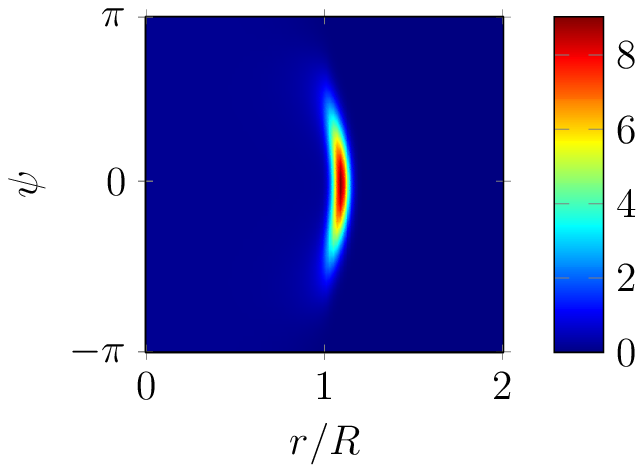}%
}\hfill
\sidesubfloat[]{%
  \label{sfig:CavityDensity_Pe=10}
  \includegraphics[width=0.275\linewidth]{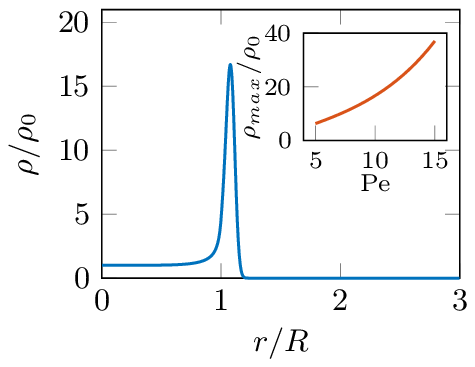}%
}\hfill
\sidesubfloat[]{%
  \label{sfig:CavityPressure_PeDependency}
  \includegraphics[width=0.29\linewidth]{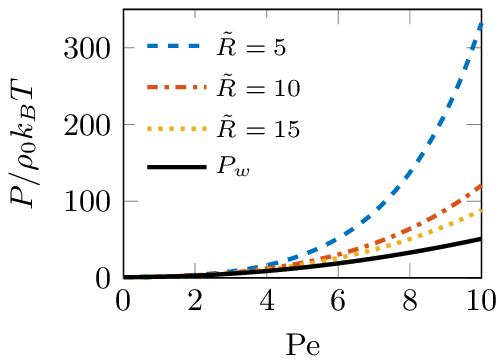}%
}\vspace{4mm}
\sidesubfloat[]{%
  \label{sfig:CavityPressure_RDependency}
  \includegraphics[width=0.29\linewidth]{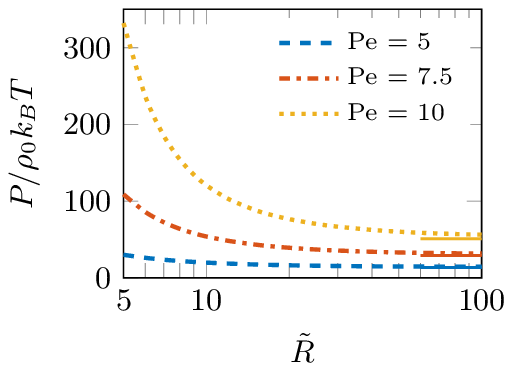}%
}\hfill
\sidesubfloat[]{%
  \label{sfig:CavityPressure_PeDependency_Chiral}
  \includegraphics[width=0.29\linewidth]{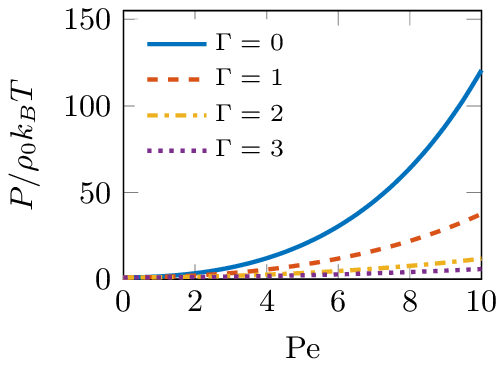}%
}\hfill
\sidesubfloat[]{%
  \label{sfig:CavityPressure_ChiralityDependency}
  \includegraphics[width=0.29\linewidth]{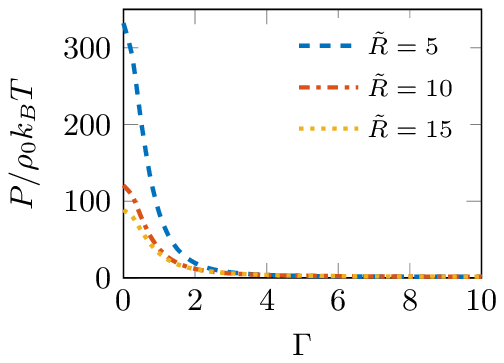}%
}
\caption{(a)-(f) Same as the panels shown in Fig. \ref{fig:object_case} but plotted here are the numerically obtained results for active particles confined in an impermeable cavity (see Fig. \ref{fig:object_case} for details and the corresponding parameter values for each panel). 
}
\label{fig:cavity_case}
\end{figure*}

\subsection{Cavity}
\label{subsec:cavity_results}

In the case of active particles confined in a circular cavity, the representative form of the numerically obtained PDF is shown in Fig.~\ref{fig:cavity_case}\subref{sfig:CavityPDF_Pe=10} for {\em non-chiral} active particles and fixed  $\Pe  = 10$ and $\tilde R=10$. The active particles form a uniform density profile in the central part of the cavity (Fig.~\ref{fig:cavity_case}\subref{sfig:CavityDensity_Pe=10}), establishing the bulk condition assumed in Section \ref{subsec:cavity_results_formulation}. The near-surface active particle orientation shows a spread around $\psi=0$, indicating that the active particles mostly orient in the normal (or radial) direction, pointing toward the boundary. The surface accumulation of active particles is much stronger here as compared with the case of a disk-shaped inclusion in a non-chiral active bath (compare maximum density values in Figs.~\ref{fig:cavity_case}\subref{sfig:CavityDensity_Pe=10} and \ref{fig:object_case}\subref{sfig:ObjectDensity_Pe=10}) and their insets). This behavior is indicative of extended detention times for active particles at a concave boundary relative to a convex (or flat) boundary \cite{Mallory2014_PRE,Smallenburg2015_PRE,Nikola2016_PRL}. There are qualitative differences between the two cases (inclusion versus cavity) as well. The maximum density of active particles at its peak near the boundary in a cavity  shows a monotonically increasing behavior with the P\'eclet number, $\Pe$, as opposed to the non-monotonic behavior found in the case of an inclusion (compare insets of Figs.~\ref{fig:cavity_case}\subref{sfig:CavityDensity_Pe=10} and \ref{fig:object_case}\subref{sfig:ObjectDensity_Pe=10}). This is because, in a cavity, the active-particle run length $\ell_{\textrm{run}}=v/D_r$ (which scales linearly with $\Pe$ in rescaled units) is bounded from above by the radius of the cavity, preventing the crossover to the ballistic regime of behavior mentioned in the case of an inclusion. The surface density of active particles within the cavity is thus larger than that found in the case of an inclusion with identical system parameter values, and so is the swim pressure as implied by Eq.~(\ref{eq:cavity_pressure_1}). 

The swim pressure on the cavity increases strongly as its radius $\tilde R$ is decreased and/or as the P\'eclet number is increased; see Figs.~\ref{fig:cavity_case}\subref{sfig:CavityPressure_PeDependency} and \subref{sfig:CavityPressure_RDependency}. The dependence on the radius of curvature is much stronger here as opposed to what we found in the case of an inclusion in Fig. \ref{fig:object_case}\subref{sfig:CavityPressure_RDependency}; in the particular example with $\Pe=10$ and  $\tilde R=5$, we find an increase about 550\% in the pressure acting on the cavity wall as compared with its reference value at a flat wall, $P_f$ (shown by black solid curve in Fig.~\ref{fig:cavity_case}\subref{sfig:CavityPressure_PeDependency}). The increase is due to the term $P_{ex}$, which is {\em positive}, contrary to the inclusion case. Our results for cavity radius smaller than the particle run length (strong confinement), $\tilde R\lesssim \Pe$, agrees with exponential decay reported in previous studies \cite{Mallory2014_Anomalous,Fily2014_Dynamics,Smallenburg2015_PRE}. 

Similar to the case of an inclusion, upon increasing the active-particle chirality, accumulation of active particles near the bounding surface of the cavity is strongly decreased (data not shown). As such, chiral active particles impart a smaller pressure on the cavity relative to non-chiral ones; see Figs.~\ref{fig:cavity_case}\subref{sfig:CavityPressure_PeDependency_Chiral} and \ref{fig:cavity_case}\subref{sfig:CavityPressure_ChiralityDependency}. The pressure decreases and tends to its equilibrium value in the case of a bulk suspension of  non-active particles as the chirality strength parameter increases to infinity. 

It is, however,  interesting to note that the pressure acting on the concave surface of a circular cavity is much more strongly dependent on active-particle chirality than the pressure acting on the convex surface of a disk-shaped inclusion with identical parameter values, as seen by comparing Figs.~\ref{fig:object_case}\subref{sfig:ObjectPressure_ChiralityDependency} and \ref{fig:cavity_case}\subref{sfig:CavityPressure_ChiralityDependency}. This property can be quantified by noting that, in both cases,  the swim pressure shows bell-shaped curves as a function of $\Gamma$ (with peaks at $\Gamma=0$). Hence, we can compare the half width at half maximum (HWHM) of the curves in the case of a disk-shaped inclusion and that of a circular cavity. Thus, in the example with $\Pe=10$ and  $\tilde R=10$, we find $\Gamma_{\mathrm{HWHM}}\simeq 1.7$ in the former case and $\Gamma_{\mathrm{HWHM}}\simeq 0.7$ in the latter case, indicating a more rapidly decaying behavior for the swim pressure with the chirality strength in the case of a cavity than in the case of an inclusion.

\floatsetup[figure]{style=plain,subcapbesideposition=top}
\begin{figure*}[t!]
\centering
\sidesubfloat[]{%
  \label{sfig:ObjectPDFChiral}
  \includegraphics[width=0.3\linewidth]{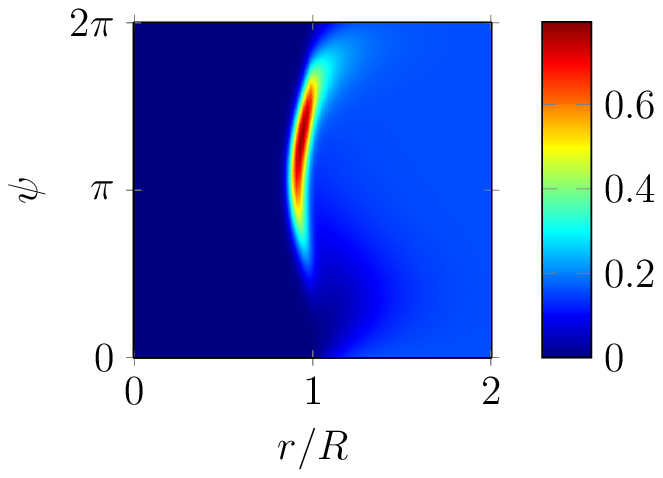}%
}\qquad
\sidesubfloat[]{%
   \label{sfig:ObjectCurrentDensity}
  \includegraphics[width=0.26\linewidth]{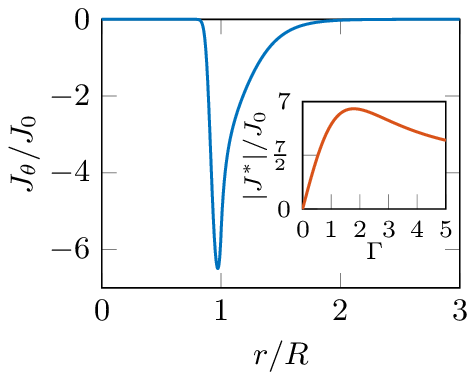}%
}
\vspace{4mm}
\sidesubfloat[]{%
  \label{sfig:CavityPDFChiral}
  \includegraphics[width=0.3\linewidth]{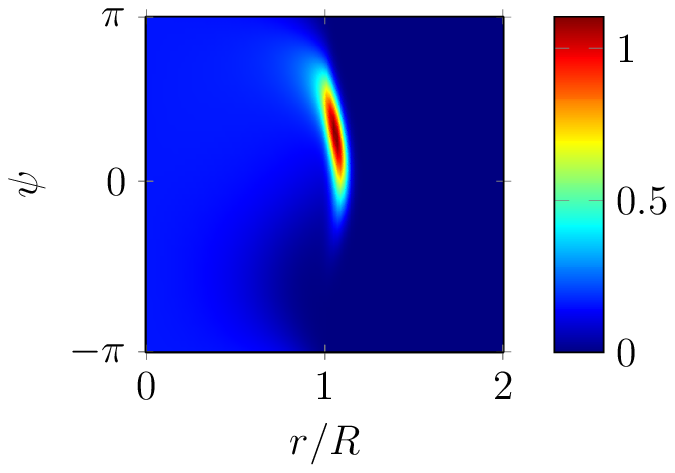}%
}\qquad
\sidesubfloat[]{%
  \label{sfig:CavityCurrentDensity}
  \includegraphics[width=0.26\linewidth]{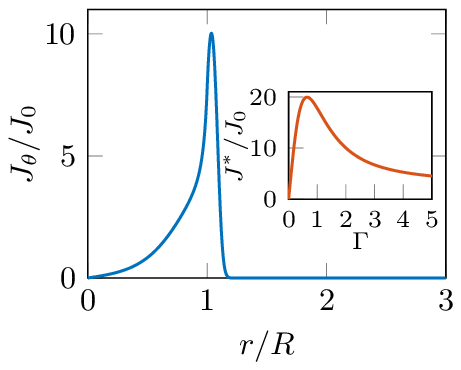}%
}
\caption{Results for chiral active particles in both cases of an inclusion (first row) and a cavity (second row). Here, we have fixed the parameter values as $\Pe = 10$, $\tilde R = 10$, and $\Gamma=2$. (a) and (c) represent the rescaled steady-state PDF, ${\cP}_0(r, \psi)/\rho_0$, of chiral particles in the coordinate plane $(r/R, \psi)$. 
(b) and (d) are the chirality-induced rotational current densities $J_\theta$ for the inclusion and cavity, respectively. The negative (positive) sign of $J_\theta$ indicates a clockwise (anti-clockwise) current near boundaries. The insets shows the influence of the chirality strength parameter $\Gamma$ on the extermum value of $J_\theta$, denoted by $J^\ast$.
}
\label{fig:current_density}
\end{figure*}

\subsection{Chirality-Induced current density}
\label{subsec:current}

As noted before, the most probable orientation of non-chiral active particles is normal to the circular boundaries in both geometries. For chiral active particle, the most probable orientation deviates from the normal-to-surface direction, creating a smaller incident angle at the surface. This gives the active particles a finite tangential velocity component on average that  produces a net rotational current next to the circular boundaries that can be evaluated from Eq. \eqref{eq:radial_current_density1}. This is accompanied by a reduction in the swim pressure (due to reduced momentum transfer to the boundaries) as discussed in detail in the preceding sections; see also Eq. \eqref{eq:pressure_current_inclusion}. 

The rescaled steady-state PDF of chiral active particles is plotted in the coordinate plane $(r/R, \psi)$ in Figs. \ref{fig:current_density}\subref{sfig:ObjectPDFChiral} and \ref{fig:current_density}\subref{sfig:CavityPDFChiral} in the  inclusion and the cavity geometries, respectively, for fixed  $\Gamma=2$, $\Pe = 10$ and $\tilde R = 10$. The PDFs show a skewed shape, which corroborate the aforementioned point regarding the shift in the most probable orientation of chiral active particles from the corresponding non-chiral values $\psi=\pi$ and $\pi=0$ in the case of inclusion and cavity (coinciding with  $\psi\simeq 4.2$ and  $\psi\simeq 0.8$ in Figs. \ref{fig:current_density}\subref{sfig:ObjectPDFChiral} and \ref{fig:current_density}\subref{sfig:CavityPDFChiral}),  respectively. 

The current density profiles, rescaled by the characteristic current $J_0=\rho_0\sqrt{D_t D_r}$, are shown in Figs.~\ref{fig:current_density}\subref{sfig:ObjectCurrentDensity} and \ref{fig:current_density}\subref{sfig:CavityCurrentDensity}.  The rotational current takes a sizable magnitude near the bounding surface; it is negative, or clockwise, in the case of an inclusion and positive, or anti-clockwise, in the case of a cavity. These particular signs for the resulting currents are due to our choice of anti-clockwise particle chirality ($\Gamma>0$) in the plots; they will be reversed if one reverses the chirality sign.  
 
Intuitively, the emergence of a net, rotational, surface current in the system can be understood by recalling that chiral active particle traverse circular (even though noisy) trajectories. Far away from the boundaries, the overlapping anti-clockwise/clockwise rotation of particles that traverse through many such trajectories, passing through a given reference point in space, lead to a vanishing net particle current density at that point. The situation is different when the reference point is taken in the vicinity of a bounding surface as the region inside (outside) the inclusion (cavity) is impermeable to active particles. As such, a statistically significant fraction of particle trajectories that go through the given reference point (out of  the whole ensemble of trajectories permitted in the absence of the bounding surface) are excluded, giving rise to an uncompensated, chirality-induced, rotational current near the surface. 

The current densities vanish in two different limits of vanishing chirality ($\Gamma=0$), as mentioned before, and infinite chirality ($\Gamma\rightarrow \infty$), as the latter case coincides with a system of non-active particles (Appendix \ref{sec:app_effective_Smoluchowski_Eq}), chosen to be in equilibrium, where macroscopic currents identically vanish. This implies that a non-monotonic behavior for surface current density (e.g., the absolute  current density at its maximum, $|J^\ast|/J_0$) as a function of $\Gamma$, as indeed seen in the insets of Figs.~\ref{fig:current_density}\subref{sfig:ObjectCurrentDensity} and \ref{fig:current_density}\subref{sfig:CavityCurrentDensity}. It is interesting to note that, for a given set of parameters, the same magnitude for the maximum current density in a cavity can be achieved by taking a comparatively smaller chirality strength as compared with the situation in the case of an inclusion; also, the rescaled current density magnitude is much larger in the former case, signifying the role of the concave boundaries in enhancing chirality-induced surface currents of active particles. 

\section{Young-Laplace equation}
\label{sec:Young_Laplace} 
 
Our results for the swim pressure on convex and concave boundaries can be used to predict the mechanical equilibrium of fluid enclosures suspended in a bulk fluid, either or both of which may contain active particles. Examples of such fluid enclosures (which we shall generically refer to as `droplets') may include  micro-compartments such as vesicles, lipid domains, immiscible or stabilized droplets in emulsions, or the recently studied realizations of active droplets \cite{Cates_PNAS2010,Tjhung_PNAS2012,Tjhung_PNAS2016,Vladescu_PRL2014,Paoluzzi_SR2016,Hennes_PNAS2017,Jin_2018,Julicher:Nature2016,Julicher_PRE2015,Julicher_NJP2017,Golestanian_Nature2017,Hyman_2014,Cate_2018review,Israelachvili2015,deGennes_Capillarity_2004,Ostwald1900}. Our goal is to explore possible consequences of the swim pressure within a first-step model, providing an active-fluid analogue for the celebrated Young-Laplace equation. The latter is standardly used to describe the capillary behavior of droplets in contact with a surrounding bulk,  both of which contain normal fluids (such as simple molecular fluids or non-active colloidal mixtures). In the case of interest here with a circular region (drop) of radius $R$, with inside and outside pressures $P_{in}$ and $P_{out}$, respectively, the Young-Laplace equation gives the pressure jump (or, the Laplace pressure, $\Delta P$) across the interface separating the two fluid regions as  \cite{deGennes_Capillarity_2004,Israelachvili2015}
\begin{equation}
\label{eq:Young-Laplace}
 \Delta P \equiv P_{in} - P_{out} = \frac{\gamma}{R}, 
\end{equation}
where $\gamma$ is the interfacial tension. Thus, in the case of normal fluids, the inside pressure of a suspended droplet is found to increase because of its interfacial tension as the droplet radius is decreased. When two such fluid droplets come in contact or are otherwise interconnected, the internal fluid flows from the smaller droplet (with higher inside pressure) to the larger one (with lower inside pressure), making the former shrink and eventually disappear in favor of the latter; a paradigm usually referred to as Ostwald ripening \cite{Israelachvili2015,deGennes_Capillarity_2004,Ostwald1900}.  

As we shall see, this general picture does not necessarily hold when the system contains active particles. In what follows, and for a clearer demonstration of the underlying mechanisms, we shall consider only the case where active particles are present outside the drops, being filled themselves by a normal fluid. In fact, the reverse scenario in which active particles are confined inside the droplets qualitatively follows the same standard paradigm as discussed above. This is because, taking the droplets to be in local mechanical equilibrium with their surrounding normal bulk fluid, which is held itself at a fixed external hydrostatic pressure $P_{out}^{(0)}$, the behavior of the inside droplet pressure with the droplet radius trivially accords with that of the interfacial tension term. This will however not be the case in the former scenario. 

In order to proceed, we adopt the simple model of a circular impermeable enclosure, whose elasticity is described by the interfacial tension $\gamma$, and calculate the swim pressure using the same methods as described in Sections \ref{sec:swimm_pressure} and \ref{sec:results}.

\subsection{Standard versus anomalous capillarity}
\label{subsec:capillarity}

\begin{figure}[t!]
\centering
\sidesubfloat[]{%
 \label{sfig:Pressure_YL}
 \includegraphics[width=0.55\linewidth]{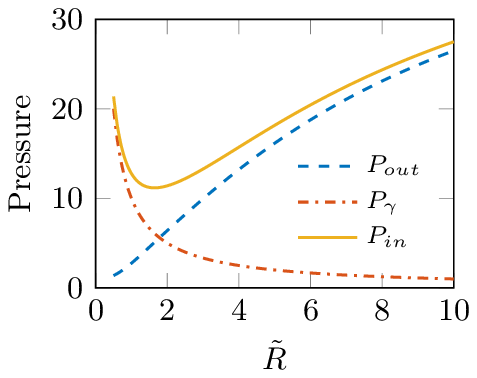}%
}\\
\vspace{4mm}
\sidesubfloat[]{%
 \label{sfig:Min_Radius}
 \includegraphics[width=0.55\linewidth]{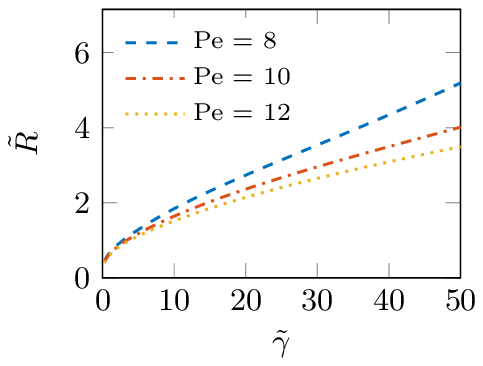}%
}
\caption{(a) The outside pressure, $ P_{out}$, the interfacial pressure, ${P}_\gamma$, and the inside pressure, $ P_{in}$,  in rescaled units as a function of the rescaled droplet radius, $\tilde{R}$, for fixed $\tilde \gamma =10$ and $\Pe=10$. The droplet is devoid of active particles, which are non-chiral and present only in the surrounding bulk. (b) The threshold radius $\tilde{R}^\ast$ plotted as a function of $\tilde{\gamma}$ for different values of $\Pe$, as indicated on the graph.}  
\label{fig:Young_Laplace_0}
\end{figure}

Let us first focus on the case of a normal-fluid droplet suspended in a bulk fluid containing {\em non-chiral} active particles with $\Gamma=0$. The external surface of the droplet experiences an excess swim pressure due to the active particles. With no loss of generality, we discard the external reference pressure  $P_{out}^{(0)}$ and use $P_{out}$ to denote the outside swim pressure, which will itself be a non-trivial function of the droplet radius. In Fig.~\ref{fig:Young_Laplace_0}\subref{sfig:Pressure_YL}, we show the outside pressure, $ P_{out}$, the interfacial pressure, ${P}_\gamma =\gamma/{R}$, and the resulting inside pressure, $ P_{in}$,  determined through  Eq.~(\ref{eq:Young-Laplace}), as a function of the rescaled droplet radius for fixed dimensionless parameter values $\tilde \gamma \equiv \gamma/(a\rho_0k_B T)=10$ and $\Pe=10$ (in all cases below, the pressure components are rescaled by  $\rho_0 k_B T$). 

In the absence of active particles, the $R$-dependence of the inside pressure follows that of the interfacial tension by decaying monotonically and inversely with the droplet radius (red dot-dashed curve). In the presence of active particles, the outside pressure shows a non-trivial dependence on the droplet radius; one with an opposing trend of monotonically increasing with $\tilde R$ (blue dashed line; see also Fig. \ref{fig:object_case}\subref{sfig:ObjectPressure_RDependency}). The resulting inside pressure can thus become a non-monotonic function of the droplet radius, exhibiting a minimum value at a finite radius (denoted by $\tilde{R}^\ast$), as seen in Fig.~\ref{fig:Young_Laplace_0}\subref{sfig:Pressure_YL} (orange solid curve). 

One can thus distinguish two different regimes of behavior for a drop, which contains a normal fluid and is suspended in an active bulk fluid: For $\tilde{R}<\tilde{R}^\ast$, the inside pressure increases as the droplet radius is decreased, representing the {\em standard capillary regime}, while for  $\tilde{R}>\tilde{R}^\ast$, the inside pressure shows the inverse and counterintuitive trend of increasing with the droplet radius, representing the {\em anomalous capillary regime}. The latter is a direct consequence of the dominant outside swim pressure. The threshold radius depends on $\Pe$ and $\tilde{\gamma}$, which can be varied to map out the behavior of the system across the parameter space. Figure~\ref{fig:Young_Laplace_0}\subref{sfig:Min_Radius} gives a representative $(\tilde R, \tilde \gamma)$-plot for three different values of $\Pe$, with the plotted curve in each case showing $\tilde{R}^\ast = \tilde{R}^\ast(\Pe, \tilde{\gamma})$, i.e., the boundary between the standard and anomalous capillary regimes  (lower and upper regions, respectively). 

\floatsetup[figure]{style=plain,subcapbesideposition=top}

\floatsetup[figure]{style=plain,subcapbesideposition=top}
\begin{figure*}[t!]
\hspace{-5cm}
\begin{minipage}[t]{0.35\textwidth}\begin{center}
\sidesubfloat[]{%
 \label{sfig:TwoDrops_1}
 \includegraphics[width=0.7\linewidth]{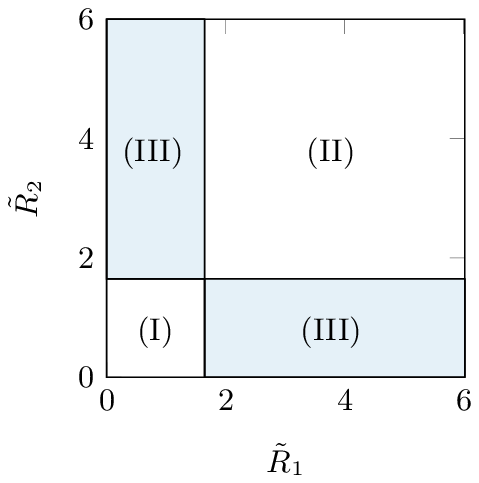}%
}\\
\vspace{-6mm}
\sidesubfloat[]{%
 \label{sfig:TwoDrops_2}
 \includegraphics[width=0.7\linewidth]{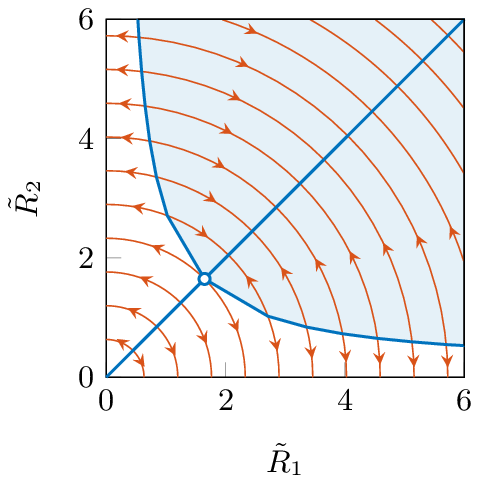}%
}
\end{center}\end{minipage}
\begin{minipage}[t]{0.4\textwidth}\begin{center}
\sidesubfloat[]{%
 \label{sfig:TwoDrops_1b}
 \includegraphics[width=0.67\linewidth]{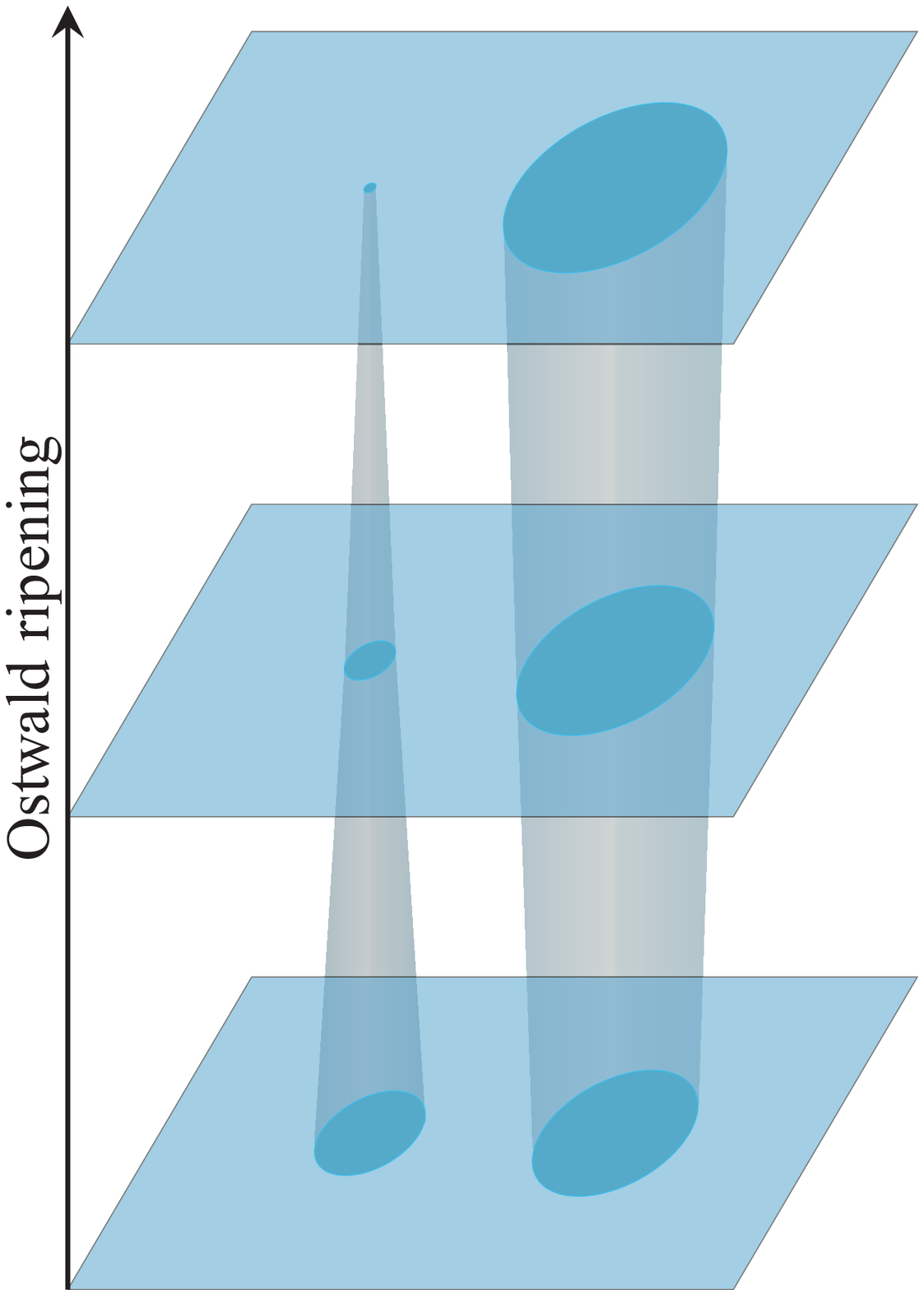}%
}\,\,\,\,\,~\sidesubfloat[]{%
 \label{sfig:TwoDrops_2b}
 \includegraphics[width=0.67\linewidth]{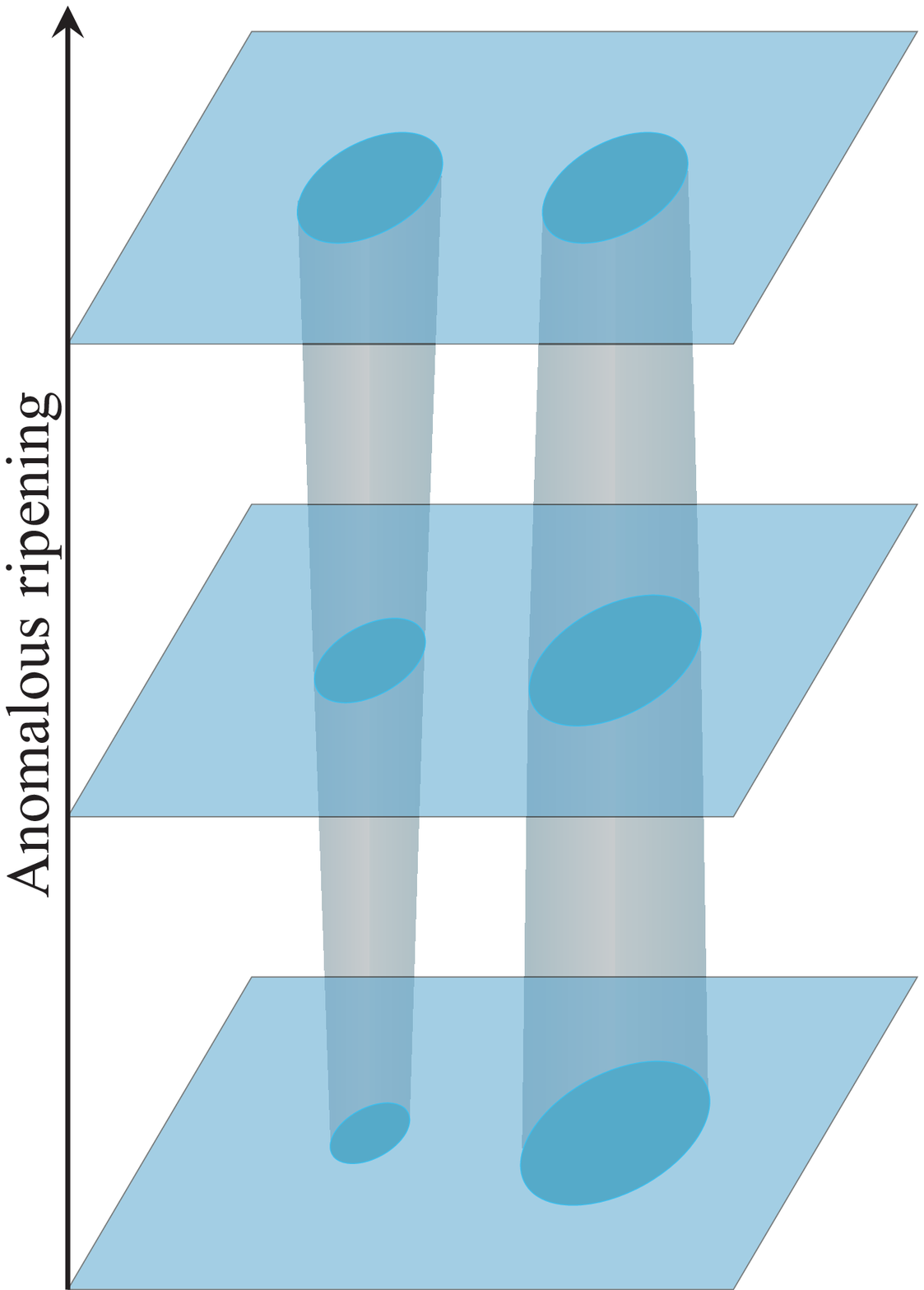}%
}
\end{center}\end{minipage}
\caption{(a) The $(\tilde{R}_1,\tilde{R}_2)$-plane is conveniently divided into three rectangular regions (I) for $\tilde R_1, \tilde R_2<\tilde R^\ast$, (II) for $\tilde R_1, \tilde R_2>\tilde R^\ast$, and (III) for $\tilde R_1>\tilde R^\ast$ and $\tilde R_2<\tilde R^\ast$ and vice versa, where $\tilde R^\ast$ is the rescaled threshold droplet radius. (b) The construction in panel (a) is used to obtain the regimes of Ostwald ripening (white region) and anomalous droplet ripening (light blue region) for two interconnected droplets of a normal incompressible fluid suspended in an active bulk. The blue straight line and the blue hyperbolic-shaped curve show the states of mechanical equilibrium, whose stability is indicated by the arrowheads marked over orange circular arcs. The latter give trajectories of evolution for the radii of two unequal-sized droplets upon interconnection toward or away from the mentioned mechanical equilibrium states; see the text for details. The open circle shows the threshold point $(\tilde{R}^\ast,\tilde{R}^\ast)$.  In panel (b), we have fixed  the parameters as $\Pe =10$, $\tilde\gamma =10$ and $\Gamma=0$, in which case we have $R^\ast \simeq 1.65$. Schematic pictures in panels (c) and (d) show the Ostwald ripening and the anomalous ripening behaviors of normal-fluid droplets suspended in an active bulk. The direction of arrows shows the size evolution from an initial state (bottom plate) toward the final state (top plate), resembling the process depicted in Refs. \cite{Julicher:Nature2016,Golestanian_Nature2017} for chemically active droplets. 
}
\label{fig:Young_Laplace_1}
\end{figure*}

\subsection{Anomalous droplet ripening}
\label{subsec:Ostwald}

The above findings predict an intriguing range of behaviors for normal-fluid droplets in an active bulk. In particular  example of two droplets of initial rescaled radii $\tilde R_1$ and $\tilde R_2$ in mechanical equilibrium with the surrounding bulk fluid, we can again predict two different regimes of standard and anomalous  behavior for the fate of the droplets after they are interconnected. To identify these regimes, and for the sake of presentation, we first divide the $(\tilde{R}_1,\tilde{R}_2)$-plane into three rectangular regions: Region (I), where both droplet radii are smaller than the threshold radius $\tilde R_1, \tilde R_2<\tilde R^\ast$; region (II), where the radii are both larger than the threshold $\tilde R_1, \tilde R_2>\tilde R^\ast$; and region III, where one is larger and the other is smaller than the threshold radius; see Fig.~\ref{fig:Young_Laplace_1}\subref{sfig:TwoDrops_1}. The condition of mechanical equilibrium implies that, upon interconnecting the two drops, the internal fluid flows from the droplet with higher inside pressure to the one with lower inside pressure. We further assume that the droplets contain an incompressible (normal) fluid of the same type and fixed total amount, giving  $\tilde{R}_1^2+\tilde{R}_2^2={\textrm{const}}$. This constrains the evolution of droplet radii to occur over circular arcs of fixed radius in the $(\tilde{R}_1,\tilde{R}_2)$-plane (orange solid curves in Fig.~\ref{fig:Young_Laplace_1}\subref{sfig:TwoDrops_2}). 

In regions (I) and (II), the states of mechanical equilibrium occur only when the interconnected droplets have also equal radii of curvature as the two droplets either exhibit standard or anomalous capillarity. This is because the inside pressure of a droplet is a monotonic function of its radius in each of these regions (see Fig.~\ref{fig:Young_Laplace_0}\subref{sfig:Pressure_YL}). The mentioned states thus fall over the line $\tilde R_2=\tilde R_1$, shown by the blue solid line bisecting the plane in Fig.~\ref{fig:Young_Laplace_1}\subref{sfig:TwoDrops_2}. 

In region (I), the line of mechanical equilibrium (bisector) is unstable. The arrowheads in the plot are used to indicate the direction of the evolution of the states that are positioned off the bisector (representing two droplets with $\tilde R_2\neq \tilde R_1<\tilde{R}^\ast$) toward the horizontal or the vertical axis (representing shrinkage of the smaller droplet and inflation of the larger one to its maximum admissible size). This behavior qualitatively agrees with the {\em Ostwald ripening} for two interconnected droplets of unequal size in normal fluids (see the schematic picture in Fig. \ref{fig:Young_Laplace_1}\subref{sfig:TwoDrops_1b}). It indicates that the swim pressure is subdominant in the parameter region (I), as also corroborated by the data in Figs.~\ref{fig:Young_Laplace_0}\subref{sfig:Pressure_YL} and \subref{sfig:Min_Radius}. 

In region (II), we find the reverse behavior: The bisector represents states of stable mechanical equilibrium and, as such, acts as the line `attractor' for all other states in this region (directions of evolution are again indicated by the arrowheads). This means that the internal fluid flows from the larger droplet to the smaller one upon interconnection, bringing them to a final state of equal size! This behavior will be referred to as {\em anomalous droplet ripening} (Fig. \ref{fig:Young_Laplace_1}\subref{sfig:TwoDrops_2b}), in which the swim pressure acting on the external, convex, surfaces of the droplets is the dominant underlying factor. 

In region (III), the situation is somewhat more intricate as the interplay between the interfacial tension and the outside swim pressure can lead to both types of ripening behaviors. In this region, the smaller droplet falls within the standard capillary regime, while the larger one falls within the anomalous capillary regime, for which the interfacial tension and the swim pressure are the most dominant factors, respectively (see Section \ref{subsec:capillarity}). Because of the non-monotonic form of the pressure profile, the state of mechanical equilibrium can still occur, even though the two interconnected droplets are of unequal sizes. These peculiar states of mechanical equilibrium can be obtained from the corresponding pressure profile in Fig.~\ref{fig:Young_Laplace_0}\subref{sfig:Pressure_YL} and are shown in the $(\tilde{R}_1,\tilde{R}_2)$-plane by the blue, hyperbolic-shaped, curve. They turn out to be unstable (this includes the threshold point shown by the open circle): The states {\em below} the mentioned curve move away from it toward the horizontal or the vertical axis, representing an Ostwald-ripening behavior, while the states {\em above} it  move toward the bisector in region (II), representing an anomalous droplet-ripening behavior. This can be understood by noting that if the smaller droplet has a higher (lower) inside pressure, i.e., the initial state of the two unequal-sized droplets is a point on the left (right) side of the upper branch of the hyperbolic-shaped curve in the $(\tilde{R}_1,\tilde{R}_2)$-plane, then the smaller droplet shrinks (inflates) due to fluid outflow (inflow) after the two droplets are interconnected. 

\begin{figure}[t!]
\centering
 \includegraphics[width=0.55\linewidth]{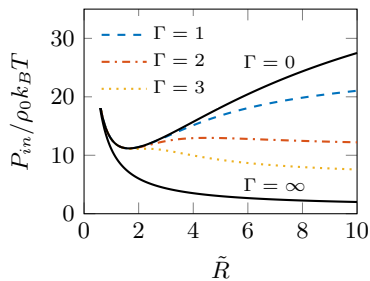}
\caption{The inside pressure, $ P_{in}$,  of a normal-fluid droplet in an active bulk as a function of the rescaled droplet radius, $\tilde R$, for fixed $\tilde \gamma =10$, $\Pe=10$, and different chirality strengths in the surrounding bulk, as shown on the graph.}
\label{fig:Young_Laplace_2}
\end{figure}

So far, we have assumed that the active particles in the outside fluid are non-chiral. Our results for the inside pressure of a droplet containing normal fluid in a chiral active bulk are shown in Fig. \ref{fig:Young_Laplace_2} for different particle chirality strengths. They show that, while the non-monotonic behavior of the inside pressure remains intact for sufficiently small droplet radii (within a range approximately set by the criterion $\tilde R\lesssim \tilde R_\omega$), chirality effects come into play and significantly reduce the inside pressure at large droplet radii. As expected, the non-monotonic behavior of the inside pressure profile is washed out when chirality effects dominate and $\tilde R_\omega$ becomes smaller than the threshold $\tilde{R}^\ast$. Hence, varying the chirality strength engenders a crossover between the two regimes of standard and anomalous capillary phenomena, which we will not delve into any further here.

\floatsetup[figure]{style=plain,subcapbesideposition=top}

\section{Summary and Discussion}
\label{sec:conclusion}

We investigate the swim pressure exerted by non-chiral and chiral self-propelled particles on convex or concave circular  boundaries, representing a fixed  inclusion immersed in an active bath and a cavity enclosing the active particles. The circular boundaries are also used to study the mechanical equilibria of suspended fluid enclosures (or, `droplets') in an active or normal fluid. The active particles are studied within a first-step model of non-interacting  Brownian particles, self-propelling at a constant linear speed, and generally also at a given chirality (intrinsic angular velocity), as commonly considered in their two-dimensional Langevin formulation  \cite{bechinger_review}. 

We provide an extensive analysis of the role of the radius of curvature, the P\'eclet number and chirality  of active particles on their spatial distribution and the swim pressure they exert on the bounding surfaces. The dependence of the swim pressure on the aforementioned system parameters are shown to be much  more dramatic, when active particles are considered next to a concave interface than to a convex interface. In the case of chiral particles, we also calculate the chirality-induced rotational current of active particles around the circular boundaries in both cases. The current  is shown to be directly related to the chirality-induced reduction in the swim pressure. It exhibits a non-monotonic behavior with the chirality strength as it vanishes in both  limits of zero and infinite chirality strength. We provide a systematic proof for the well-known results that particle chirality suppresses self-propulsion effects  (see, e.g., Refs. \cite{Xue:EPJST2014,Xue:EPL2015,Mahdi2017_ChiralSwimmers}). 
    
As an interesting application of our results, we then study the mechanical equilibrium of fluid droplets characterized by an interfacial tension and show that the standard capillary paradigms, described by the Young-Laplace equation (in, e.g.,  their classic manifestation of Ostwald ripening), may not be applicable in the case of active fluids. This is particularly true when suspended droplets of normal fluid are considered within a bulk fluid containing active particles, in which case the outside pressure acting on the external surface of the droplets admits a non-trivial dependence on the droplet size because of the swim pressure of active particles. This is shown to lead to a non-monotonic dependence of the inside droplet pressure on the droplet radius: While sufficiently small droplets in an active bulk behave in accordance with the standard capillary paradigm, i.e., the inside droplet pressure decreases with increasing its size, the larger droplets do not and their inside pressure can increase with the droplet size. This {\em anomalous capillary} behavior becomes dominant at sufficiently large P\'eclet number of active particles, sufficiently large droplet radius (larger than a given threshold) and also at sufficiently small interfacial tension. It is most pronounced when active particles are non-chiral; this is because chiral particles produce smaller outside swim pressures and, as such, the mentioned anomaly is suppressed as the particles become increasingly more chiral.  

In the textbook example of two interconnected droplets of different sizes, we identify two characteristically different regimes of behavior for the fate of the droplets. In the standard capillary regime, mechanical equilibrium occurs when the two droplets have equal sizes and thus equal inside pressures. Also, the smaller droplet will have a higher inside pressure as compared with the larger one. Hence, the smaller droplet shrinks as the larger one grows in size after they are interconnected, representing the Ostwald ripening effect. In the anomalous capillary regime, mechanical equilibrium (with equal inside pressure for the two interconnected drops) is permitted not only when the two droplets have equal sizes, but also when the two droplets are of unequal sizes. The latter gives a novel curve of mechanical equilibria across the parameter space, which is however shown to be unstable, while the former type of mechanical equilibria (characterized by equal droplet radii) define a stable line `attractor'  in the parameter space. Consequently,  we identify a regime of {\em anomalous droplet ripening}, where the  unequal sizes of two interconnected droplets change until they reach equal values. 

Our results signify the potential role of active particles in drastically modifying the capillary equilibria of suspended droplets in active bulk  fluids. These predictions yield themselves rather easily to experimental verification as active particles need to be introduced merely through the bulk fluid, which can itself be prepared to include a suspension of stabilized droplets (or simply two normal-fluid vesicles/capsules with manual interconnection), whose favored eventual state is thus predicted to be one of mono-disperse droplets. 

Possible extensions of our analysis include modeling capillary dynamics of droplets \cite{Brennen_2013}, and studying the role of swim pressure in the stability of large emulsions, where fluid  droplets may be mobile and interact also through the effective interactions mediated by their surrounding active particles \cite{Mahdi2017_ChiralSwimmers}. Our current model is based on a few simplifying assumption for the active particles, which can be improved by accounting for steric and/or  hydrodynamic interactions (see, e.g., Refs. \cite{Hernandez-Ortiz1,wallattraction,li2011accumulation,wallattraction3,Underhill_2014,ardekani,Schaar2015_PRL,Mathijssen:2016c}), inter-particle alignment effects (see, e.g., Refs. \cite{ramaswamyreview,Marchetti:RMP2013,gompper_review} and references therein),  as well as the role of active noise (represented by fluctuating self-propulsion and angular velocities) in the dynamics of active particles \cite{Romanczuk:EPJ2012}. 

\section{Conflicts of interest}

There are no conflicts of interest to declare. 

\section{Acknowledgements}
 
A.N. acknowledges partial support from Iran Science Elites Federation and the Associateship Scheme of The Abdus Salam International Centre for Theoretical Physics (Trieste, Italy). We thank M. Mohammadi, M. Sebtosheikh and M.N. Popescu for useful comments. 

\section{Appendices}
\appendix 

\section{Analytical results for swim pressure}
\label{sec:app_Pressure_Equation}

Integrating Eq.~(\ref{eq:master_eq_polar}) over the angle $\psi$ from $0$ to $2\pi$ gives 
\begin{equation}
\label{eq:first_integral_eq}
vm_1-\mu_t \rho\,\partial_r V-D_t\partial_r \rho =0,
\end{equation}
which relates the zeroth-order moment of the PDF, which is the number density $\rho(r)$, and the first-order moment, $m_1(r)$ (note that the above relation is equivalent to $J_r=0$ with $J_r$ defined in Appendix \ref{sec:app_Current_Density}). Using Eq. \eqref{eq:first_integral_eq} and \eqref{eq:pressure_inclusion}, one can immediately obtain Eqs. \eqref{eq:pressure_1} and \eqref{eq:cavity_pressure_1} for the cases of an inclusion and a cavity, respectively. 

Repeating the aforementioned operation by first  multiplying Eq.~(\ref{eq:master_eq_polar})  by $\cos\psi$, we find
\begin{align}
\label{eq:second integral eq}
\nonumber
&\partial_r\left[r\bigg(\frac{v}{2}(\rho+m_2)-\mu_tm_1\partial_r V- D_t \partial_r m_1\bigg)\right] \\ \nonumber
&-\frac{v}{2}(\rho-m_2)+ r \omega \int_0^{2\pi}  \sin\psi\,\cP_0(r,\psi)\,{\mathrm{d}}\psi \\
&+\frac{1}{r}(D_t+D_r r^2) m_1 =0.
\end{align}
It is worth-mentioning that $m_1(r)$ and $m_2(r)$ are negligible everywhere unless in a narrow vicinity of the interfacial region $r\simeq R$ \cite{footnote2}. For sufficiently large radius of curvature, the relatively small $D_t/r$ term in the third line of Eq.~(\ref{eq:second integral eq}) can safely be neglected, yielding 
\begin{align}
\label{eq:second_moment_approx}
\nonumber
D_r m_1&\simeq \frac{1}{r} \big(-vm_2+\mu_t m_1 \partial_r V+D_t \partial_r m_1 \big) \\ \nonumber
&- \partial_r\left[\frac{v}{2}(\rho+m_2)-\mu_tm_1\partial_r V- D_t \partial_r m_1\right] \\
&- \omega \int_0^{2\pi}  \sin\psi\,\cP_0\,{\mathrm{d}}\psi.
\end{align}
Since $m_1, \partial_r m_1$, and $m_2$ vanish at the origin and in the bulk, and by integrating Eq. \eqref{eq:second_moment_approx} over $r$ in the case of an inclusion, we find
\begin{equation}
\label{eq:second_moment_integral}
D_r \int_\Lambda^0 m_1(r)\,{\mathrm{d}}r \simeq \cI(\Lambda) + \frac{v}{2}\rho_0 - \omega \int_0^{2\pi} \!\!\!\int_\Lambda^0  \sin\psi\,\cP_0\,{\mathrm{d}}\psi\,{\mathrm{d}}r,
\end{equation}
where $\rho_0=\rho(\Lambda)$ is the bulk density and
\begin{equation}
\cI(\Lambda) = \int_\Lambda^0  \big(-vm_2+\mu_t m_1 \partial_r V+D_t \partial_r m_1 \big)\frac{ {\mathrm{d}}r}{r}.
\end{equation}
Inserting Eq.~(\ref{eq:second_moment_integral}) into Eq.~(\ref{eq:pressure_1}) gives Eqs. \eqref{eq:P_decomp}-\eqref{eq:pressure_inclusion_2c} for the pressure in the inclusion case. The derivation in the case of a cavity is similar to the one outlined above  with the only difference that the integrals over $r$ needs to be taken over the interval  $[0, \infty)$ (see Section \ref{subsec:cavity_results_formulation}). 

\section{Chirality-induced current}
\label{sec:app_Current_Density}

The steady-state current density of active particles can be obtained by integrating Eq. (\ref{eq:master_eq_polar}) over the orientational degree of freedom, $\varphi$,  and writing the result as $\del\cdot\bJ = 0$; $\bJ$ is  the desired current density that can be expressed in polar coordinates as $\bJ = J_r \be_r + J_\theta \be_\theta$, where
\begin{align}
\label{eq:J_components}
J_r = \int_0^{2\pi}\! \cA_r(\br,\varphi) \,{\mathrm{d}}\varphi,\quad J_\theta = \int_0^{2\pi}\! \cA_\theta(\br,\varphi) \,{\mathrm{d}}\varphi,
\end{align}
and  
\begin{align}
\label{eq:A_r}
\cA_r &= [v\cos(\varphi-\theta)  - \mu_t \partial_r V]\cP_0-D_t\partial_r\cP_0 , \\ 
\label{eq:A_theta}
\cA_\theta &= v\sin(\varphi-\theta) \cP_0-\frac{D_t}{r}\partial_\theta\cP_0.
\end{align}
On general grounds, one can argue $J_r = 0$ (this can be proved by applying the divergence theorem for $\bJ$ in a circular domain of arbitrary radius centered at the origin and by noting that, due to rotational symmetry,  $\bJ$ depends only on  $r$). This result holds regardless of the specific choice of system parameters. The rotational current $J_\theta$ can take nonzero values, except in the non-chiral case, $\Gamma=0$, where  $J_\theta = 0$. Using the expressions from Eqs.~(\ref{eq:J_components}) and (\ref{eq:A_theta}) (and by changing the angular integration variable from $\varphi$ to $\psi$), we arrive at Eq. \eqref{eq:radial_current_density1}. 

\section{Numerical scheme}
\label{sec:app_num_scheme}

We first cast Eq. (\ref{eq:master_eq_polar}) in the non-dimensionalized form by rescaling $r$  as $\tilde r=r/a$ and the potential $V(r)$ as $\tilde{V}(\tilde r)=V(a\tilde r)/(k_{\mathrm{B}}T)$, obtaining 
\begin{equation}
\label{eq:dimensionless_div_eq}
 \tilde{\del}_\psi\cdot\tilde{\bcJ} = \frac{1}{\tilde r}  \partial_{\tilde r}(\tilde r \tilde{\cJ}_{\tilde r}) +  \frac{1}{\tilde r} \partial_\psi\tilde{\cJ}_\psi=0
\end{equation}
with
\begin{align}
\tilde{\cJ}_{\tilde r} &=\big(  \Pe \cos\psi- \partial_{\tilde r} \tilde{V} \big)\tilde\cP_0-  \partial_{\tilde r} \tilde\cP_0, \\
\tilde{\cJ}_\psi &= (- \Pe \sin\psi + \tilde r\,\Gamma) \,\tilde\cP_0 - \frac{1}{\tilde r}( 1 + {\tilde r}^2) \partial_\psi \tilde\cP_0, 
\end{align} 
where $\tilde \cP_0(\tilde r, \psi)  =  \cP_0(a\tilde r, \psi)/\rho_0$ is the rescaled steady-state PDF, which is to be calculated by solving Eq.~(\ref{eq:dimensionless_div_eq}) numerically. This is done using COMSOL Multiphysics v5.2. In the case of an inclusion, the equation is solved over the rectangular domain specified by $\psi\in[0,2\pi)$ and $\tilde r\in[0,\tilde{r}_m]$, where $\tilde{r}_m$ must be large enough to ensure the bulk condition of constant density. By numerical inspection, and for the range of parameters considered in this work, we find that $\tilde{r}_m=3\tilde R$ is an appropriate, numerically efficient choice. The cyclic nature of $\psi$ entails the condition $\tilde\cP_0(\tilde r,\psi+2\pi)=\tilde\cP_0(\tilde r,\psi)$. We also use Dirichlet boundary condition as $\tilde\cP_0(\tilde{r}_m,\psi)=1$ (being an arbitrary choice of probability normalization). In the case of a cavity, and for the sake of presentation, we take the computational domain slightly differently as $\psi\in[-\pi,\pi)$  and $\tilde r\in[0,\tilde{r}_m]$ and assume $\tilde{r}_m=3\tilde R$, periodic boundary condition along $\psi$ direction and the Dirichlet boundary condition $\tilde\cP_0(0,\psi)=1$. 

\section{Limit of large chirality -- Adiabatic elimination of fast variables}
\label{sec:app_effective_Smoluchowski_Eq}

As noted in the text, the limiting behavior of a system of active self-propelled particles, when the chirality strength becomes large, coincides with that of a corresponding system of non-active (passive) particles. Here, we give a systematic proof of the validity of this statement using the standard methods known as adiabatic elimination of fast variables \cite{Risken_FP_1996}.

Let us first note that the magnitude of the chirality strength parameter, $|\Gamma|=|\omega|/D_r$, is nothing but the ratio of the characteristic timescale of rotational diffusion, $D_r^{-1}$, and the  intrinsic (active) angular velocity of the particles, $|\omega|^{-1}$. Therefore, $|\Gamma|\gg 1$ represents a situation, where the chirality timescale is much shorter than the other timescales in the system. Such a separation of timescales enables an adiabatic elimination  \cite{Risken_FP_1996} of fast variables (i.e., the angular degree of freedom), leading to an {\em effective} governing equation for the PDF of the slow variables (i.e., the translational degrees of freedom).  This can be achieved through a perturbative scheme analogous to the Born-Oppenheimer approximation \cite{Risken_FP_1996} by writing the Smoluchowski-Fokker-Planck equation (\ref{eq:master_eq}) as
\begin{equation}
\label{eq:Smoluchowski_eq_2nd_form}
\partial_t\cP(\br,\varphi, t) = \big(\cL + \omega \cM \big) \cP(\br,\varphi,t),
\end{equation} 
in which $\cL$ and $\cM$ are the operators
\begin{align}
\label{eq:L_op}
\cL &= \mu_t \del^2 V - (v\bu  - \mu_t \del V)\cdot\del \nonumber\\
&\quad +D_t\del^2 + D_r\partial_\varphi^2,\\
\cM &= -\partial_\varphi.
\end{align}
We expand the time-dependent PDF $\cP(\br,\varphi,t)$ in terms of the eigenfunctions, $\{f_n(\varphi)\}$, of the operator $\cM$ as follows
\begin{equation}
\label{eq:PDF_expansion}
 \cP(\br,\varphi,t) = \sum_{n=-\infty}^\infty c_n(\br,t) f_n(\varphi). 
\end{equation}
The eigenfunctions of $\cM$ follow immediately from the equation $\cM f_n(\varphi) = \lambda_n f_n(\varphi)$, giving the discrete set of normalized eigenfunctions $f_n=\exp(-\lambda_n\varphi)/\sqrt{2\pi}$ defined over the domain $\varphi\in[0,2\pi)$. Since $\varphi$ is a cyclic variable, the eigenvalues are $\lambda_n = \dot\iota n$ for $n\in\mathbb{Z}$. These eigenfunctions form a complete orthonormal set
\begin{align}
\int_0^{2\pi} &f_n^\ast(\varphi) f_m(\varphi) \, {\mathrm{d}}\varphi = \delta_{nm}, \\
\sum_{n=-\infty}^\infty &f_n^\ast(\varphi) f_n(\varphi^\prime) = \delta(\varphi-\varphi^\prime).
\end{align}
Inserting Eq. \eqref{eq:PDF_expansion} into Eq.~(\ref{eq:Smoluchowski_eq_2nd_form}) and multiplying both sides of Eq.~(\ref{eq:Smoluchowski_eq_2nd_form}) by $f_n^\ast$ from the left and integrating over $\varphi$, we find
\begin{equation}
\label{eq:c_t}
\big( \partial_t - \omega \lambda_n \big) c_n = \sum_{m=-\infty}^\infty \cL_{nm} c_m,
\end{equation}
where 
\begin{equation}
 \cL_{nm} = \int_0^{2\pi} f_n^\ast(\varphi) \cL f_m(\varphi) \, {\mathrm{d}}\varphi.
\end{equation}
Since we are interested in the regime of very large $\omega$, we can neglect the time derivative in Eq. \eqref{eq:c_t} except for $n=0$, in which case  $\lambda_0 =0$; therefore, 
\begin{align}
\label{eq:c_0_t}
\partial_t c_0 &= \sum_{m=-\infty}^\infty \cL_{0m} c_m, 
\end{align}
while for $n \neq 0$, we have 
\begin{align}
\nonumber
c_n &= -(\omega \lambda_n)^{-1} \sum_{m=-\infty}^\infty \cL_{nm} c_m, \\ 
&= -(\omega \lambda_n)^{-1} \cL_{n0}\,c_0 + O(\omega^{-2}), 
\label{eq:c_n_t}
\end{align}
where we have also used the fact that $c_0\sim{\mathcal O}(1)$ and $c_n\sim{\mathcal O}(\omega^{-1})$.  Using Eqs.~(\ref{eq:c_0_t}) and (\ref{eq:c_n_t}), we have 
\begin{equation}
\label{eq:c_0_t_final}
 \partial_t c_0 = \cL_{00}\,c_0 - \omega^{-1} \left( \sum_{m\neq 0} \cL_{0m} \lambda_m^{-1} \cL_{m0} \right)\,c_0 + O(\omega^{-2}).
\end{equation}
It is easy to verify that $c_0=c_0(\br,t)$ is proportional to the time-dependent PDF of the slow variables, $\cP(\br,t) = \int_0^{2\pi} \cP(\br,\varphi,t)\,{\mathrm{d}}\varphi$, that is, $c_0= \cP(\br,t)/\sqrt{2\pi}$. Equation \eqref{eq:c_0_t_final} thus represents the effective equation governing the PDF of slow variables, $\br$. To the leading order, or by taking the limit $|\omega| \rightarrow \infty$ (or, $|\Gamma|\rightarrow\infty$), we find 
\begin{equation}
\label{eq:effective_FP_slow_variables}
 \partial_t \cP(\br,t) = \cL_{00}\,\cP(\br,t), 
\end{equation}
where $\cL_{00} = \int_0^{2\pi} f_0^\ast \cL f_0 \, {\mathrm{d}}\varphi$, which, using Eq. \eqref{eq:L_op} and  $f_0=1/\sqrt{2\pi}$, follows  as
\begin{equation}
 \cL_{00} = \mu_t \del^2 V + \mu_t \del V\cdot\del + D_t\del^2.
\end{equation}
The final equation \eqref{eq:effective_FP_slow_variables} is therefore nothing but the Smoluchowski-Fokker-Planck of the corresponding non-active (passive) system, when active self-propulsion is switched off (in which case, the rotational diffusion also plays no roles). This proves that, assuming the Einstein-Smoluchowski-Sutherland relation, the steady-state behavior of the system in the infinite chirality limit coincides with the equilibrium behavior. 

It is interesting to note that the same conclusion as above can be reached, when rotational diffusion is taken as the fastest process (or, when $D_r$ is taken to infinity). In this case, the operator $\cM$ can be chosen as $\cM = \partial_\varphi^2$. The same procedure as outlined above can be repeated, leading to the non-active governing equation \eqref{eq:effective_FP_slow_variables}.  

\bibliographystyle{aipnum4-1}

\bibliography{Refs}

\begin{thebibliography}{99}%
\makeatletter
\providecommand \@ifxundefined [1]{%
 \@ifx{#1\undefined}
}%
\providecommand \@ifnum [1]{%
 \ifnum #1\expandafter \@firstoftwo
 \else \expandafter \@secondoftwo
 \fi
}%
\providecommand \@ifx [1]{%
 \ifx #1\expandafter \@firstoftwo
 \else \expandafter \@secondoftwo
 \fi
}%
\providecommand \natexlab [1]{#1}%
\providecommand \enquote  [1]{``#1''}%
\providecommand \bibnamefont  [1]{#1}%
\providecommand \bibfnamefont [1]{#1}%
\providecommand \citenamefont [1]{#1}%
\providecommand \href@noop [0]{\@secondoftwo}%
\providecommand \href [0]{\begingroup \@sanitize@url \@href}%
\providecommand \@href[1]{\@@startlink{#1}\@@href}%
\providecommand \@@href[1]{\endgroup#1\@@endlink}%
\providecommand \@sanitize@url [0]{\catcode `\\12\catcode `\$12\catcode
  `\&12\catcode `\#12\catcode `\^12\catcode `\_12\catcode `\%12\relax}%
\providecommand \@@startlink[1]{}%
\providecommand \@@endlink[0]{}%
\providecommand \url  [0]{\begingroup\@sanitize@url \@url }%
\providecommand \@url [1]{\endgroup\@href {#1}{\urlprefix }}%
\providecommand \urlprefix  [0]{URL }%
\providecommand \Eprint [0]{\href }%
\providecommand \doibase [0]{http://dx.doi.org/}%
\providecommand \selectlanguage [0]{\@gobble}%
\providecommand \bibinfo  [0]{\@secondoftwo}%
\providecommand \bibfield  [0]{\@secondoftwo}%
\providecommand \translation [1]{[#1]}%
\providecommand \BibitemOpen [0]{}%
\providecommand \bibitemStop [0]{}%
\providecommand \bibitemNoStop [0]{.\EOS\space}%
\providecommand \EOS [0]{\spacefactor3000\relax}%
\providecommand \BibitemShut  [1]{\csname bibitem#1\endcsname}%
\let\auto@bib@innerbib\@empty
\bibitem [{\citenamefont {Lauga}\ and\ \citenamefont
  {Powers}(2009)}]{Lauga:RPP009}%
  \BibitemOpen
  \bibfield  {author} {\bibinfo {author} {\bibfnamefont {E.}~\bibnamefont
  {Lauga}}\ and\ \bibinfo {author} {\bibfnamefont {T.~R.}\ \bibnamefont
  {Powers}},\ }\href {https://doi.org/10.1088/0034-4885/72/9/096601} {\bibfield
   {journal} {\bibinfo  {journal} {Rep. Prog. Phys.}\ }\textbf {\bibinfo
  {volume} {72}},\ \bibinfo {pages} {096601} (\bibinfo {year}
  {2009})}\BibitemShut {NoStop}%
\bibitem [{\citenamefont {Ramaswamy}(2010)}]{ramaswamyreview}%
  \BibitemOpen
  \bibfield  {author} {\bibinfo {author} {\bibfnamefont {S.}~\bibnamefont
  {Ramaswamy}},\ }\href
  {https://doi.org/10.1146/annurev-conmatphys-070909-104101} {\bibfield
  {journal} {\bibinfo  {journal} {Annu. Rev. Condens. Matter Phys.}\ }\textbf
  {\bibinfo {volume} {1}},\ \bibinfo {pages} {323} (\bibinfo {year}
  {2010})}\BibitemShut {NoStop}%
\bibitem [{\citenamefont {Golestanian}, \citenamefont {Yeomans},\ and\
  \citenamefont {Uchida}(2011)}]{golestanian_review}%
  \BibitemOpen
  \bibfield  {author} {\bibinfo {author} {\bibfnamefont {R.}~\bibnamefont
  {Golestanian}}, \bibinfo {author} {\bibfnamefont {J.~M.}\ \bibnamefont
  {Yeomans}}, \ and\ \bibinfo {author} {\bibfnamefont {N.}~\bibnamefont
  {Uchida}},\ }\href {https://doi.org/10.1039/c0sm01121e} {\bibfield  {journal}
  {\bibinfo  {journal} {Soft Matter}\ }\textbf {\bibinfo {volume} {7}},\
  \bibinfo {pages} {3074} (\bibinfo {year} {2011})}\BibitemShut {NoStop}%
\bibitem [{\citenamefont {Romanczuk}\ \emph {et~al.}(2012)\citenamefont
  {Romanczuk}, \citenamefont {B{\"a}r}, \citenamefont {Ebeling}, \citenamefont
  {Lindner},\ and\ \citenamefont {Schimansky-Geier}}]{Romanczuk:EPJ2012}%
  \BibitemOpen
  \bibfield  {author} {\bibinfo {author} {\bibfnamefont {P.}~\bibnamefont
  {Romanczuk}}, \bibinfo {author} {\bibfnamefont {M.}~\bibnamefont {B{\"a}r}},
  \bibinfo {author} {\bibfnamefont {W.}~\bibnamefont {Ebeling}}, \bibinfo
  {author} {\bibfnamefont {B.}~\bibnamefont {Lindner}}, \ and\ \bibinfo
  {author} {\bibfnamefont {L.}~\bibnamefont {Schimansky-Geier}},\ }\href
  {https://doi.org/10.1140/epjst/e2012-01529-y} {\bibfield  {journal} {\bibinfo
   {journal} {Eur. Phys. J. Spec. Top.}\ }\textbf {\bibinfo {volume} {202}},\
  \bibinfo {pages} {1} (\bibinfo {year} {2012})}\BibitemShut {NoStop}%
\bibitem [{\citenamefont {Marchetti}\ \emph {et~al.}(2013)\citenamefont
  {Marchetti}, \citenamefont {Joanny}, \citenamefont {Ramaswamy}, \citenamefont
  {Liverpool}, \citenamefont {Prost}, \citenamefont {Rao},\ and\ \citenamefont
  {Simha}}]{Marchetti:RMP2013}%
  \BibitemOpen
  \bibfield  {author} {\bibinfo {author} {\bibfnamefont {M.~C.}\ \bibnamefont
  {Marchetti}}, \bibinfo {author} {\bibfnamefont {J.~F.}\ \bibnamefont
  {Joanny}}, \bibinfo {author} {\bibfnamefont {S.}~\bibnamefont {Ramaswamy}},
  \bibinfo {author} {\bibfnamefont {T.~B.}\ \bibnamefont {Liverpool}}, \bibinfo
  {author} {\bibfnamefont {J.}~\bibnamefont {Prost}}, \bibinfo {author}
  {\bibfnamefont {M.}~\bibnamefont {Rao}}, \ and\ \bibinfo {author}
  {\bibfnamefont {R.~A.}\ \bibnamefont {Simha}},\ }\href
  {https://doi.org/10.1103/RevModPhys.85.1143} {\bibfield  {journal} {\bibinfo
  {journal} {Rev. Mod. Phys.}\ }\textbf {\bibinfo {volume} {85}},\ \bibinfo
  {pages} {1143} (\bibinfo {year} {2013})}\BibitemShut {NoStop}%
\bibitem [{\citenamefont {Yeomans}, \citenamefont {Pushkin},\ and\
  \citenamefont {Shum}(2014)}]{Yeomans:EPJ2014}%
  \BibitemOpen
  \bibfield  {author} {\bibinfo {author} {\bibfnamefont {J.~M.}\ \bibnamefont
  {Yeomans}}, \bibinfo {author} {\bibfnamefont {D.~O.}\ \bibnamefont
  {Pushkin}}, \ and\ \bibinfo {author} {\bibfnamefont {H.}~\bibnamefont
  {Shum}},\ }\href {https://doi.org/10.1140/epjst/e2014-02225-8} {\bibfield
  {journal} {\bibinfo  {journal} {Eur. Phys. J. Spec. Top.}\ }\textbf {\bibinfo
  {volume} {223}},\ \bibinfo {pages} {1771} (\bibinfo {year}
  {2014})}\BibitemShut {NoStop}%
\bibitem [{\citenamefont {Elgeti}, \citenamefont {Winkler},\ and\ \citenamefont
  {Gompper}(2015)}]{gompper_review}%
  \BibitemOpen
  \bibfield  {author} {\bibinfo {author} {\bibfnamefont {J.}~\bibnamefont
  {Elgeti}}, \bibinfo {author} {\bibfnamefont {R.~G.}\ \bibnamefont {Winkler}},
  \ and\ \bibinfo {author} {\bibfnamefont {G.}~\bibnamefont {Gompper}},\ }\href
  {https://doi.org/10.1088/0034-4885/78/5/056601} {\bibfield  {journal}
  {\bibinfo  {journal} {Rep. Prog. Phys.}\ }\textbf {\bibinfo {volume} {78}},\
  \bibinfo {pages} {056601} (\bibinfo {year} {2015})}\BibitemShut {NoStop}%
\bibitem [{\citenamefont {Bechinger}\ \emph {et~al.}(2016)\citenamefont
  {Bechinger}, \citenamefont {{Di Leonardo}}, \citenamefont {L{\"o}wen},
  \citenamefont {Reichhardt}, \citenamefont {Volpe},\ and\ \citenamefont
  {Volpe}}]{bechinger_review}%
  \BibitemOpen
  \bibfield  {author} {\bibinfo {author} {\bibfnamefont {C.}~\bibnamefont
  {Bechinger}}, \bibinfo {author} {\bibfnamefont {R.}~\bibnamefont {{Di
  Leonardo}}}, \bibinfo {author} {\bibfnamefont {H.}~\bibnamefont {L{\"o}wen}},
  \bibinfo {author} {\bibfnamefont {C.}~\bibnamefont {Reichhardt}}, \bibinfo
  {author} {\bibfnamefont {G.}~\bibnamefont {Volpe}}, \ and\ \bibinfo {author}
  {\bibfnamefont {G.}~\bibnamefont {Volpe}},\ }\href
  {https://doi.org/10.1103/RevModPhys.88.045006} {\bibfield  {journal}
  {\bibinfo  {journal} {Rev. Mod. Phys.}\ }\textbf {\bibinfo {volume} {88}},\
  \bibinfo {pages} {045006} (\bibinfo {year} {2016})}\BibitemShut {NoStop}%
\bibitem [{\citenamefont {Z{\"o}ttl}\ and\ \citenamefont
  {Stark}(2016)}]{ZottlStark_review}%
  \BibitemOpen
  \bibfield  {author} {\bibinfo {author} {\bibfnamefont {A.}~\bibnamefont
  {Z{\"o}ttl}}\ and\ \bibinfo {author} {\bibfnamefont {H.}~\bibnamefont
  {Stark}},\ }\href {https://doi.org/10.1088/0953-8984/28/25/253001} {\bibfield
   {journal} {\bibinfo  {journal} {J. Phys.: Condens. Matter}\ }\textbf
  {\bibinfo {volume} {28}},\ \bibinfo {pages} {253001} (\bibinfo {year}
  {2016})}\BibitemShut {NoStop}%
\bibitem [{\citenamefont {Paxton}\ \emph {et~al.}(2006)\citenamefont {Paxton},
  \citenamefont {Sundararajan}, \citenamefont {Mallouk},\ and\ \citenamefont
  {Sen}}]{Paxton2006_review}%
  \BibitemOpen
  \bibfield  {author} {\bibinfo {author} {\bibfnamefont {W.~F.}\ \bibnamefont
  {Paxton}}, \bibinfo {author} {\bibfnamefont {S.}~\bibnamefont
  {Sundararajan}}, \bibinfo {author} {\bibfnamefont {T.~E.}\ \bibnamefont
  {Mallouk}}, \ and\ \bibinfo {author} {\bibfnamefont {A.}~\bibnamefont
  {Sen}},\ }\href {https://doi.org/10.1002/anie.200600060} {\bibfield
  {journal} {\bibinfo  {journal} {Angew. Chem.}\ }\textbf {\bibinfo {volume}
  {45}},\ \bibinfo {pages} {5420} (\bibinfo {year} {2006})}\BibitemShut
  {NoStop}%
\bibitem [{\citenamefont {Koumakis}\ \emph {et~al.}(2013)\citenamefont
  {Koumakis}, \citenamefont {Lepore}, \citenamefont {Maggi},\ and\
  \citenamefont {{Di Leonardo}}}]{cargo}%
  \BibitemOpen
  \bibfield  {author} {\bibinfo {author} {\bibfnamefont {N.}~\bibnamefont
  {Koumakis}}, \bibinfo {author} {\bibfnamefont {A.}~\bibnamefont {Lepore}},
  \bibinfo {author} {\bibfnamefont {C.}~\bibnamefont {Maggi}}, \ and\ \bibinfo
  {author} {\bibfnamefont {R.}~\bibnamefont {{Di Leonardo}}},\ }\href
  {https://doi.org/10.1038/ncomms3588} {\bibfield  {journal} {\bibinfo
  {journal} {Nat. Commun.}\ }\textbf {\bibinfo {volume} {4}},\ \bibinfo {pages}
  {2588} (\bibinfo {year} {2013})}\BibitemShut {NoStop}%
\bibitem [{\citenamefont {Cheang}\ \emph {et~al.}(2014)\citenamefont {Cheang},
  \citenamefont {Lee}, \citenamefont {Julius},\ and\ \citenamefont
  {Kim}}]{robotic}%
  \BibitemOpen
  \bibfield  {author} {\bibinfo {author} {\bibfnamefont {U.~K.}\ \bibnamefont
  {Cheang}}, \bibinfo {author} {\bibfnamefont {K.}~\bibnamefont {Lee}},
  \bibinfo {author} {\bibfnamefont {A.~A.}\ \bibnamefont {Julius}}, \ and\
  \bibinfo {author} {\bibfnamefont {M.~J.}\ \bibnamefont {Kim}},\ }\href
  {https://doi.org/10.1063/1.4893695} {\bibfield  {journal} {\bibinfo
  {journal} {Appl. Phys. Lett.}\ }\textbf {\bibinfo {volume} {105}},\ \bibinfo
  {pages} {083705} (\bibinfo {year} {2014})}\BibitemShut {NoStop}%
\bibitem [{\citenamefont {Medina-S{\'a}nchez}\ \emph
  {et~al.}(2016)\citenamefont {Medina-S{\'a}nchez}, \citenamefont {Schwarz},
  \citenamefont {Meyer}, \citenamefont {Hebenstreit},\ and\ \citenamefont
  {Schmidt}}]{sperm-carrying}%
  \BibitemOpen
  \bibfield  {author} {\bibinfo {author} {\bibfnamefont {M.}~\bibnamefont
  {Medina-S{\'a}nchez}}, \bibinfo {author} {\bibfnamefont {L.}~\bibnamefont
  {Schwarz}}, \bibinfo {author} {\bibfnamefont {A.~K.}\ \bibnamefont {Meyer}},
  \bibinfo {author} {\bibfnamefont {F.}~\bibnamefont {Hebenstreit}}, \ and\
  \bibinfo {author} {\bibfnamefont {O.~G.}\ \bibnamefont {Schmidt}},\ }\href
  {https://doi.org/10.1021/acs.nanolett.5b04221} {\bibfield  {journal}
  {\bibinfo  {journal} {Nano Lett.}\ }\textbf {\bibinfo {volume} {16}},\
  \bibinfo {pages} {555} (\bibinfo {year} {2016})}\BibitemShut {NoStop}%
\bibitem [{\citenamefont {Walther}\ and\ \citenamefont
  {M{\"u}ller}(2013)}]{janusmain}%
  \BibitemOpen
  \bibfield  {author} {\bibinfo {author} {\bibfnamefont {A.}~\bibnamefont
  {Walther}}\ and\ \bibinfo {author} {\bibfnamefont {A.~H.~E.}\ \bibnamefont
  {M{\"u}ller}},\ }\href {https://doi.org/10.1021/cr300089t} {\bibfield
  {journal} {\bibinfo  {journal} {Chem. Rev.}\ }\textbf {\bibinfo {volume}
  {113}},\ \bibinfo {pages} {5194} (\bibinfo {year} {2013})}\BibitemShut
  {NoStop}%
\bibitem [{\citenamefont {Ebbens}\ and\ \citenamefont
  {Howse}(2010)}]{Ebbens2010_Pursuit}%
  \BibitemOpen
  \bibfield  {author} {\bibinfo {author} {\bibfnamefont {S.~J.}\ \bibnamefont
  {Ebbens}}\ and\ \bibinfo {author} {\bibfnamefont {J.~R.}\ \bibnamefont
  {Howse}},\ }\href {https://doi.org/10.1039/B918598D} {\bibfield  {journal}
  {\bibinfo  {journal} {Soft Matter}\ }\textbf {\bibinfo {volume} {6}},\
  \bibinfo {pages} {726} (\bibinfo {year} {2010})}\BibitemShut {NoStop}%
\bibitem [{\citenamefont {Howse}\ \emph {et~al.}(2007)\citenamefont {Howse},
  \citenamefont {Jones}, \citenamefont {Ryan}, \citenamefont {Gough},
  \citenamefont {Vafabakhsh},\ and\ \citenamefont
  {Golestanian}}]{Howse2007_Self-Motile}%
  \BibitemOpen
  \bibfield  {author} {\bibinfo {author} {\bibfnamefont {J.~R.}\ \bibnamefont
  {Howse}}, \bibinfo {author} {\bibfnamefont {R.~A.~L.}\ \bibnamefont {Jones}},
  \bibinfo {author} {\bibfnamefont {A.~J.}\ \bibnamefont {Ryan}}, \bibinfo
  {author} {\bibfnamefont {T.}~\bibnamefont {Gough}}, \bibinfo {author}
  {\bibfnamefont {R.}~\bibnamefont {Vafabakhsh}}, \ and\ \bibinfo {author}
  {\bibfnamefont {R.}~\bibnamefont {Golestanian}},\ }\href
  {https://doi.org/10.1103/PhysRevLett.99.048102} {\bibfield  {journal}
  {\bibinfo  {journal} {Phys. Rev. Lett.}\ }\textbf {\bibinfo {volume} {99}},\
  \bibinfo {pages} {048102} (\bibinfo {year} {2007})}\BibitemShut {NoStop}%
\bibitem [{\citenamefont {Goldstein}(2015)}]{goldstein_review}%
  \BibitemOpen
  \bibfield  {author} {\bibinfo {author} {\bibfnamefont {R.~E.}\ \bibnamefont
  {Goldstein}},\ }\href {https://doi.org/10.1146/annurev-fluid-010313-141426}
  {\bibfield  {journal} {\bibinfo  {journal} {Annu. Rev. Fluid Mech.}\ }\textbf
  {\bibinfo {volume} {47}},\ \bibinfo {pages} {343} (\bibinfo {year}
  {2015})}\BibitemShut {NoStop}%
\bibitem [{\citenamefont {Lauga}(2016)}]{Lauga:ANNREVF2016}%
  \BibitemOpen
  \bibfield  {author} {\bibinfo {author} {\bibfnamefont {E.}~\bibnamefont
  {Lauga}},\ }\href {https://doi.org/10.1146/annurev-fluid-122414-034606}
  {\bibfield  {journal} {\bibinfo  {journal} {Annu. Rev. Fluid Mech.}\ }\textbf
  {\bibinfo {volume} {48}},\ \bibinfo {pages} {105} (\bibinfo {year}
  {2016})}\BibitemShut {NoStop}%
\bibitem [{\citenamefont {Elgeti}\ and\ \citenamefont
  {Gompper}(2016)}]{Elgeti2016_Microswimmers}%
  \BibitemOpen
  \bibfield  {author} {\bibinfo {author} {\bibfnamefont {J.}~\bibnamefont
  {Elgeti}}\ and\ \bibinfo {author} {\bibfnamefont {G.}~\bibnamefont
  {Gompper}},\ }\href {https://doi.org/10.1140/epjst/e2016-60070-6} {\bibfield
  {journal} {\bibinfo  {journal} {Eur. Phys. J. Spec. Top.}\ }\textbf {\bibinfo
  {volume} {225}},\ \bibinfo {pages} {2333} (\bibinfo {year}
  {2016})}\BibitemShut {NoStop}%
\bibitem [{\citenamefont {Elgeti}\ and\ \citenamefont
  {Gompper}(2013)}]{Elgeti2013_WallAccumulation}%
  \BibitemOpen
  \bibfield  {author} {\bibinfo {author} {\bibfnamefont {J.}~\bibnamefont
  {Elgeti}}\ and\ \bibinfo {author} {\bibfnamefont {G.}~\bibnamefont
  {Gompper}},\ }\href {https://doi.org/10.1209/0295-5075/101/48003} {\bibfield
  {journal} {\bibinfo  {journal} {EPL (Europhysics Letters)}\ }\textbf
  {\bibinfo {volume} {101}},\ \bibinfo {pages} {48003} (\bibinfo {year}
  {2013})}\BibitemShut {NoStop}%
\bibitem [{\citenamefont {Rusconi}, \citenamefont {Guasto},\ and\ \citenamefont
  {Stocker}(2014)}]{rusconi}%
  \BibitemOpen
  \bibfield  {author} {\bibinfo {author} {\bibfnamefont {R.}~\bibnamefont
  {Rusconi}}, \bibinfo {author} {\bibfnamefont {J.~S.}\ \bibnamefont {Guasto}},
  \ and\ \bibinfo {author} {\bibfnamefont {R.}~\bibnamefont {Stocker}},\ }\href
  {https://doi.org/10.1038/nphys2883} {\bibfield  {journal} {\bibinfo
  {journal} {Nat. Phys.}\ }\textbf {\bibinfo {volume} {10}},\ \bibinfo {pages}
  {212} (\bibinfo {year} {2014})}\BibitemShut {NoStop}%
\bibitem [{\citenamefont {Hernandez-Ortiz}, \citenamefont {Stoltz},\ and\
  \citenamefont {Graham}(2005)}]{Hernandez-Ortiz1}%
  \BibitemOpen
  \bibfield  {author} {\bibinfo {author} {\bibfnamefont {J.~P.}\ \bibnamefont
  {Hernandez-Ortiz}}, \bibinfo {author} {\bibfnamefont {C.~G.}\ \bibnamefont
  {Stoltz}}, \ and\ \bibinfo {author} {\bibfnamefont {M.~D.}\ \bibnamefont
  {Graham}},\ }\href {https://doi.org/10.1103/PhysRevLett.95.204501} {\bibfield
   {journal} {\bibinfo  {journal} {Phys. Rev. Lett.}\ }\textbf {\bibinfo
  {volume} {95}},\ \bibinfo {pages} {204501} (\bibinfo {year}
  {2005})}\BibitemShut {NoStop}%
\bibitem [{\citenamefont {Berke}\ \emph {et~al.}(2008)\citenamefont {Berke},
  \citenamefont {Turner}, \citenamefont {Berg},\ and\ \citenamefont
  {Lauga}}]{wallattraction}%
  \BibitemOpen
  \bibfield  {author} {\bibinfo {author} {\bibfnamefont {A.~P.}\ \bibnamefont
  {Berke}}, \bibinfo {author} {\bibfnamefont {L.}~\bibnamefont {Turner}},
  \bibinfo {author} {\bibfnamefont {H.~C.}\ \bibnamefont {Berg}}, \ and\
  \bibinfo {author} {\bibfnamefont {E.}~\bibnamefont {Lauga}},\ }\href
  {https://doi.org/10.1103/PhysRevLett.101.038102} {\bibfield  {journal}
  {\bibinfo  {journal} {Phys. Rev. Lett.}\ }\textbf {\bibinfo {volume} {101}},\
  \bibinfo {pages} {038102} (\bibinfo {year} {2008})}\BibitemShut {NoStop}%
\bibitem [{\citenamefont {Li}\ \emph {et~al.}(2011)\citenamefont {Li},
  \citenamefont {Bensson}, \citenamefont {Nisimova}, \citenamefont {Munger},
  \citenamefont {Mahautmr}, \citenamefont {Tang}, \citenamefont {Maxey},\ and\
  \citenamefont {Brun}}]{li2011accumulation}%
  \BibitemOpen
  \bibfield  {author} {\bibinfo {author} {\bibfnamefont {G.}~\bibnamefont
  {Li}}, \bibinfo {author} {\bibfnamefont {J.}~\bibnamefont {Bensson}},
  \bibinfo {author} {\bibfnamefont {L.}~\bibnamefont {Nisimova}}, \bibinfo
  {author} {\bibfnamefont {D.}~\bibnamefont {Munger}}, \bibinfo {author}
  {\bibfnamefont {P.}~\bibnamefont {Mahautmr}}, \bibinfo {author}
  {\bibfnamefont {J.~X.}\ \bibnamefont {Tang}}, \bibinfo {author}
  {\bibfnamefont {M.~R.}\ \bibnamefont {Maxey}}, \ and\ \bibinfo {author}
  {\bibfnamefont {Y.~V.}\ \bibnamefont {Brun}},\ }\href
  {https://doi.org/10.1103/PhysRevE.84.041932} {\bibfield  {journal} {\bibinfo
  {journal} {Phys. Rev. E}\ }\textbf {\bibinfo {volume} {84}},\ \bibinfo
  {pages} {041932} (\bibinfo {year} {2011})}\BibitemShut {NoStop}%
\bibitem [{\citenamefont {Spagnolie}\ and\ \citenamefont
  {Lauga}(2012)}]{wallattraction3}%
  \BibitemOpen
  \bibfield  {author} {\bibinfo {author} {\bibfnamefont {S.~E.}\ \bibnamefont
  {Spagnolie}}\ and\ \bibinfo {author} {\bibfnamefont {E.}~\bibnamefont
  {Lauga}},\ }\href {https://doi.org/10.1017/jfm.2012.101} {\bibfield
  {journal} {\bibinfo  {journal} {J. Fluid Mech.}\ }\textbf {\bibinfo {volume}
  {700}},\ \bibinfo {pages} {105} (\bibinfo {year} {2012})}\BibitemShut
  {NoStop}%
\bibitem [{\citenamefont {Chilukuri}, \citenamefont {Collins},\ and\
  \citenamefont {Underhill}(2014)}]{Underhill_2014}%
  \BibitemOpen
  \bibfield  {author} {\bibinfo {author} {\bibfnamefont {S.}~\bibnamefont
  {Chilukuri}}, \bibinfo {author} {\bibfnamefont {C.~H.}\ \bibnamefont
  {Collins}}, \ and\ \bibinfo {author} {\bibfnamefont {P.~T.}\ \bibnamefont
  {Underhill}},\ }\href {https://doi.org/10.1088/0953-8984/26/11/115101}
  {\bibfield  {journal} {\bibinfo  {journal} {J. Phys.: Condens. Matter}\
  }\textbf {\bibinfo {volume} {26}},\ \bibinfo {pages} {115101} (\bibinfo
  {year} {2014})}\BibitemShut {NoStop}%
\bibitem [{\citenamefont {Li}\ and\ \citenamefont {Ardekani}(2014)}]{ardekani}%
  \BibitemOpen
  \bibfield  {author} {\bibinfo {author} {\bibfnamefont {G.-J.}\ \bibnamefont
  {Li}}\ and\ \bibinfo {author} {\bibfnamefont {A.~M.}\ \bibnamefont
  {Ardekani}},\ }\href {https://doi.org/10.1103/PhysRevE.90.013010} {\bibfield
  {journal} {\bibinfo  {journal} {Phys. Rev. E}\ }\textbf {\bibinfo {volume}
  {90}},\ \bibinfo {pages} {013010} (\bibinfo {year} {2014})}\BibitemShut
  {NoStop}%
\bibitem [{\citenamefont {Schaar}, \citenamefont {Z{\"o}ttl},\ and\
  \citenamefont {Stark}(2015)}]{Schaar2015_PRL}%
  \BibitemOpen
  \bibfield  {author} {\bibinfo {author} {\bibfnamefont {K.}~\bibnamefont
  {Schaar}}, \bibinfo {author} {\bibfnamefont {A.}~\bibnamefont {Z{\"o}ttl}}, \
  and\ \bibinfo {author} {\bibfnamefont {H.}~\bibnamefont {Stark}},\ }\href
  {https://doi.org/10.1103/PhysRevLett.115.038101} {\bibfield  {journal}
  {\bibinfo  {journal} {Phys. Rev. Lett.}\ }\textbf {\bibinfo {volume} {115}},\
  \bibinfo {pages} {038101} (\bibinfo {year} {2015})}\BibitemShut {NoStop}%
\bibitem [{\citenamefont {Mathijssen}\ \emph {et~al.}(2016)\citenamefont
  {Mathijssen}, \citenamefont {Doostmohammadi}, \citenamefont {Yeomans},\ and\
  \citenamefont {Shendruk}}]{Mathijssen:2016c}%
  \BibitemOpen
  \bibfield  {author} {\bibinfo {author} {\bibfnamefont {A.~J. T.~M.}\
  \bibnamefont {Mathijssen}}, \bibinfo {author} {\bibfnamefont
  {A.}~\bibnamefont {Doostmohammadi}}, \bibinfo {author} {\bibfnamefont
  {J.~M.}\ \bibnamefont {Yeomans}}, \ and\ \bibinfo {author} {\bibfnamefont
  {T.~N.}\ \bibnamefont {Shendruk}},\ }\href
  {https://doi.org/10.1098/rsif.2015.0936} {\bibfield  {journal} {\bibinfo
  {journal} {J. R. Soc. Interface}\ }\textbf {\bibinfo {volume} {13}} (\bibinfo
  {year} {2016})}\BibitemShut {NoStop}%
\bibitem [{\citenamefont {Lion}\ and\ \citenamefont
  {Allen}(2014)}]{Lion2014_Osmosis}%
  \BibitemOpen
  \bibfield  {author} {\bibinfo {author} {\bibfnamefont {T.~W.}\ \bibnamefont
  {Lion}}\ and\ \bibinfo {author} {\bibfnamefont {R.~J.}\ \bibnamefont
  {Allen}},\ }\href {https://doi.org/10.1209/0295-5075/106/34003} {\bibfield
  {journal} {\bibinfo  {journal} {EPL (Europhysics Letters)}\ }\textbf
  {\bibinfo {volume} {106}},\ \bibinfo {pages} {34003} (\bibinfo {year}
  {2014})}\BibitemShut {NoStop}%
\bibitem [{\citenamefont {Takatori}, \citenamefont {Yan},\ and\ \citenamefont
  {Brady}(2014)}]{Takatori2014_SwimPressure}%
  \BibitemOpen
  \bibfield  {author} {\bibinfo {author} {\bibfnamefont {S.~C.}\ \bibnamefont
  {Takatori}}, \bibinfo {author} {\bibfnamefont {W.}~\bibnamefont {Yan}}, \
  and\ \bibinfo {author} {\bibfnamefont {J.~F.}\ \bibnamefont {Brady}},\ }\href
  {https://doi.org/10.1103/PhysRevLett.113.028103} {\bibfield  {journal}
  {\bibinfo  {journal} {Phys. Rev. Lett.}\ }\textbf {\bibinfo {volume} {113}},\
  \bibinfo {pages} {028103} (\bibinfo {year} {2014})}\BibitemShut {NoStop}%
\bibitem [{\citenamefont {Yang}, \citenamefont {Manning},\ and\ \citenamefont
  {Marchetti}(2014)}]{yang_aggregation_2014}%
  \BibitemOpen
  \bibfield  {author} {\bibinfo {author} {\bibfnamefont {X.}~\bibnamefont
  {Yang}}, \bibinfo {author} {\bibfnamefont {M.~L.}\ \bibnamefont {Manning}}, \
  and\ \bibinfo {author} {\bibfnamefont {M.~C.}\ \bibnamefont {Marchetti}},\
  }\href {https://doi.org/10.1039/C4SM00927D} {\bibfield  {journal} {\bibinfo
  {journal} {Soft Matter}\ }\textbf {\bibinfo {volume} {10}},\ \bibinfo {pages}
  {6477} (\bibinfo {year} {2014})}\BibitemShut {NoStop}%
\bibitem [{\citenamefont {Fily}, \citenamefont {Baskaran},\ and\ \citenamefont
  {Hagan}(2014)}]{Fily2014_Dynamics}%
  \BibitemOpen
  \bibfield  {author} {\bibinfo {author} {\bibfnamefont {Y.}~\bibnamefont
  {Fily}}, \bibinfo {author} {\bibfnamefont {A.}~\bibnamefont {Baskaran}}, \
  and\ \bibinfo {author} {\bibfnamefont {M.~F.}\ \bibnamefont {Hagan}},\ }\href
  {https://doi.org/10.1039/C4SM00975D} {\bibfield  {journal} {\bibinfo
  {journal} {Soft Matter}\ }\textbf {\bibinfo {volume} {10}},\ \bibinfo {pages}
  {5609} (\bibinfo {year} {2014})}\BibitemShut {NoStop}%
\bibitem [{\citenamefont {Mallory}\ \emph {et~al.}(2014)\citenamefont
  {Mallory}, \citenamefont {\u{S}ari{\'c}}, \citenamefont {Valeriani},\ and\
  \citenamefont {Cacciuto}}]{Mallory2014_Anomalous}%
  \BibitemOpen
  \bibfield  {author} {\bibinfo {author} {\bibfnamefont {S.~A.}\ \bibnamefont
  {Mallory}}, \bibinfo {author} {\bibfnamefont {A.}~\bibnamefont
  {\u{S}ari{\'c}}}, \bibinfo {author} {\bibfnamefont {C.}~\bibnamefont
  {Valeriani}}, \ and\ \bibinfo {author} {\bibfnamefont {A.}~\bibnamefont
  {Cacciuto}},\ }\href {https://doi.org/10.1103/PhysRevE.89.052303} {\bibfield
  {journal} {\bibinfo  {journal} {Phys. Rev. E.}\ }\textbf {\bibinfo {volume}
  {89}},\ \bibinfo {pages} {052303} (\bibinfo {year} {2014})}\BibitemShut
  {NoStop}%
\bibitem [{\citenamefont {Mallory}, \citenamefont {Valeriani},\ and\
  \citenamefont {Cacciuto}(2014)}]{Mallory2014_PRE}%
  \BibitemOpen
  \bibfield  {author} {\bibinfo {author} {\bibfnamefont {S.~A.}\ \bibnamefont
  {Mallory}}, \bibinfo {author} {\bibfnamefont {C.}~\bibnamefont {Valeriani}},
  \ and\ \bibinfo {author} {\bibfnamefont {A.}~\bibnamefont {Cacciuto}},\
  }\href {https://doi.org/10.1103/PhysRevE.90.032309} {\bibfield  {journal}
  {\bibinfo  {journal} {Phys. Rev. E}\ }\textbf {\bibinfo {volume} {90}},\
  \bibinfo {pages} {032309} (\bibinfo {year} {2014})}\BibitemShut {NoStop}%
\bibitem [{\citenamefont {Smallenburg}\ and\ \citenamefont
  {L{\"o}wen}(2015)}]{Smallenburg2015_PRE}%
  \BibitemOpen
  \bibfield  {author} {\bibinfo {author} {\bibfnamefont {F.}~\bibnamefont
  {Smallenburg}}\ and\ \bibinfo {author} {\bibfnamefont {H.}~\bibnamefont
  {L{\"o}wen}},\ }\href {https://doi.org/10.1103/PhysRevE.92.032304} {\bibfield
   {journal} {\bibinfo  {journal} {Phys. Rev. E}\ }\textbf {\bibinfo {volume}
  {92}},\ \bibinfo {pages} {032304} (\bibinfo {year} {2015})}\BibitemShut
  {NoStop}%
\bibitem [{\citenamefont {Winkler}, \citenamefont {Wysocki},\ and\
  \citenamefont {Gompper}(2015)}]{Winkler_SM2015}%
  \BibitemOpen
  \bibfield  {author} {\bibinfo {author} {\bibfnamefont {R.~G.}\ \bibnamefont
  {Winkler}}, \bibinfo {author} {\bibfnamefont {A.}~\bibnamefont {Wysocki}}, \
  and\ \bibinfo {author} {\bibfnamefont {G.}~\bibnamefont {Gompper}},\ }\href
  {https://doi.org/10.1039/C5SM01412C} {\bibfield  {journal} {\bibinfo
  {journal} {Soft Matter}\ }\textbf {\bibinfo {volume} {11}},\ \bibinfo {pages}
  {6680} (\bibinfo {year} {2015})}\BibitemShut {NoStop}%
\bibitem [{\citenamefont {Fily}, \citenamefont {Baskaran},\ and\ \citenamefont
  {Hagan}(2015)}]{Fily2015_DynamicsDensity}%
  \BibitemOpen
  \bibfield  {author} {\bibinfo {author} {\bibfnamefont {Y.}~\bibnamefont
  {Fily}}, \bibinfo {author} {\bibfnamefont {A.}~\bibnamefont {Baskaran}}, \
  and\ \bibinfo {author} {\bibfnamefont {M.~F.}\ \bibnamefont {Hagan}},\ }\href
  {https://doi.org/10.1103/PhysRevE.91.012125} {\bibfield  {journal} {\bibinfo
  {journal} {Phys. Rev. E}\ }\textbf {\bibinfo {volume} {91}},\ \bibinfo
  {pages} {012125} (\bibinfo {year} {2015})}\BibitemShut {NoStop}%
\bibitem [{\citenamefont {Yan}\ and\ \citenamefont
  {Brady}(2015{\natexlab{a}})}]{yan_swim_2015}%
  \BibitemOpen
  \bibfield  {author} {\bibinfo {author} {\bibfnamefont {W.}~\bibnamefont
  {Yan}}\ and\ \bibinfo {author} {\bibfnamefont {J.~F.}\ \bibnamefont
  {Brady}},\ }\href {https://doi.org/10.1039/C5SM01318F} {\bibfield  {journal}
  {\bibinfo  {journal} {Soft Matter}\ }\textbf {\bibinfo {volume} {11}},\
  \bibinfo {pages} {6235} (\bibinfo {year} {2015}{\natexlab{a}})}\BibitemShut
  {NoStop}%
\bibitem [{\citenamefont {Takatori}\ and\ \citenamefont
  {Brady}(2015)}]{Takatori_PRE2015}%
  \BibitemOpen
  \bibfield  {author} {\bibinfo {author} {\bibfnamefont {S.~C.}\ \bibnamefont
  {Takatori}}\ and\ \bibinfo {author} {\bibfnamefont {J.~F.}\ \bibnamefont
  {Brady}},\ }\href {https://doi.org/10.1103/PhysRevE.91.032117} {\bibfield
  {journal} {\bibinfo  {journal} {Phys. Rev. E}\ }\textbf {\bibinfo {volume}
  {91}},\ \bibinfo {pages} {032117} (\bibinfo {year} {2015})}\BibitemShut
  {NoStop}%
\bibitem [{\citenamefont {Solon}\ \emph
  {et~al.}(2015{\natexlab{a}})\citenamefont {Solon}, \citenamefont {Fily},
  \citenamefont {Baskaran}, \citenamefont {Cates}, \citenamefont {Kafri},
  \citenamefont {Kardar},\ and\ \citenamefont {Tailleur}}]{Solon2015_Nature}%
  \BibitemOpen
  \bibfield  {author} {\bibinfo {author} {\bibfnamefont {A.~P.}\ \bibnamefont
  {Solon}}, \bibinfo {author} {\bibfnamefont {Y.}~\bibnamefont {Fily}},
  \bibinfo {author} {\bibfnamefont {A.}~\bibnamefont {Baskaran}}, \bibinfo
  {author} {\bibfnamefont {M.~E.}\ \bibnamefont {Cates}}, \bibinfo {author}
  {\bibfnamefont {Y.}~\bibnamefont {Kafri}}, \bibinfo {author} {\bibfnamefont
  {M.}~\bibnamefont {Kardar}}, \ and\ \bibinfo {author} {\bibfnamefont
  {J.}~\bibnamefont {Tailleur}},\ }\href {https://doi.org/10.1038/nphys3377}
  {\bibfield  {journal} {\bibinfo  {journal} {Nat. Phys.}\ }\textbf {\bibinfo
  {volume} {11}},\ \bibinfo {pages} {673} (\bibinfo {year}
  {2015}{\natexlab{a}})}\BibitemShut {NoStop}%
\bibitem [{\citenamefont {Solon}\ \emph
  {et~al.}(2015{\natexlab{b}})\citenamefont {Solon}, \citenamefont
  {Stenhammar}, \citenamefont {Wittkowski}, \citenamefont {Kardar},
  \citenamefont {Kafri}, \citenamefont {Cates},\ and\ \citenamefont
  {Tailleur}}]{Solon2015_PRL}%
  \BibitemOpen
  \bibfield  {author} {\bibinfo {author} {\bibfnamefont {A.~P.}\ \bibnamefont
  {Solon}}, \bibinfo {author} {\bibfnamefont {J.}~\bibnamefont {Stenhammar}},
  \bibinfo {author} {\bibfnamefont {R.}~\bibnamefont {Wittkowski}}, \bibinfo
  {author} {\bibfnamefont {M.}~\bibnamefont {Kardar}}, \bibinfo {author}
  {\bibfnamefont {Y.}~\bibnamefont {Kafri}}, \bibinfo {author} {\bibfnamefont
  {M.~E.}\ \bibnamefont {Cates}}, \ and\ \bibinfo {author} {\bibfnamefont
  {J.}~\bibnamefont {Tailleur}},\ }\href
  {https://doi.org/10.1103/PhysRevLett.114.198301} {\bibfield  {journal}
  {\bibinfo  {journal} {Phys. Rev. Lett.}\ }\textbf {\bibinfo {volume} {114}},\
  \bibinfo {pages} {198301} (\bibinfo {year} {2015}{\natexlab{b}})}\BibitemShut
  {NoStop}%
\bibitem [{\citenamefont {Yan}\ and\ \citenamefont
  {Brady}(2015{\natexlab{b}})}]{Yan_JFM2015}%
  \BibitemOpen
  \bibfield  {author} {\bibinfo {author} {\bibfnamefont {W.}~\bibnamefont
  {Yan}}\ and\ \bibinfo {author} {\bibfnamefont {J.~F.}\ \bibnamefont
  {Brady}},\ }\href {https://doi.org/10.1017/jfm.2015.621} {\bibfield
  {journal} {\bibinfo  {journal} {J. Fluid Mech.}\ }\textbf {\bibinfo {volume}
  {785}} (\bibinfo {year} {2015}{\natexlab{b}})}\BibitemShut {NoStop}%
\bibitem [{\citenamefont {Wysocki}, \citenamefont {Elgeti},\ and\ \citenamefont
  {Gompper}(2015)}]{Wysocki_2015}%
  \BibitemOpen
  \bibfield  {author} {\bibinfo {author} {\bibfnamefont {A.}~\bibnamefont
  {Wysocki}}, \bibinfo {author} {\bibfnamefont {J.}~\bibnamefont {Elgeti}}, \
  and\ \bibinfo {author} {\bibfnamefont {G.}~\bibnamefont {Gompper}},\ }\href
  {https://doi.org/10.1103/PhysRevE.91.050302} {\bibfield  {journal} {\bibinfo
  {journal} {Phys. Rev. E}\ }\textbf {\bibinfo {volume} {91}},\ \bibinfo
  {pages} {050302} (\bibinfo {year} {2015})}\BibitemShut {NoStop}%
\bibitem [{\citenamefont {Nikola}\ \emph {et~al.}(2016)\citenamefont {Nikola},
  \citenamefont {Solon}, \citenamefont {Kafri}, \citenamefont {Kardar},
  \citenamefont {Tailleur},\ and\ \citenamefont {Voituriez}}]{Nikola2016_PRL}%
  \BibitemOpen
  \bibfield  {author} {\bibinfo {author} {\bibfnamefont {N.}~\bibnamefont
  {Nikola}}, \bibinfo {author} {\bibfnamefont {A.~P.}\ \bibnamefont {Solon}},
  \bibinfo {author} {\bibfnamefont {Y.}~\bibnamefont {Kafri}}, \bibinfo
  {author} {\bibfnamefont {M.}~\bibnamefont {Kardar}}, \bibinfo {author}
  {\bibfnamefont {J.}~\bibnamefont {Tailleur}}, \ and\ \bibinfo {author}
  {\bibfnamefont {R.}~\bibnamefont {Voituriez}},\ }\href
  {https://doi.org/10.1103/PhysRevLett.117.098001} {\bibfield  {journal}
  {\bibinfo  {journal} {Phys. Rev. Lett.}\ }\textbf {\bibinfo {volume} {117}},\
  \bibinfo {pages} {098001} (\bibinfo {year} {2016})}\BibitemShut {NoStop}%
\bibitem [{\citenamefont {Ginot}\ \emph {et~al.}(2015)\citenamefont {Ginot},
  \citenamefont {Theurkauff}, \citenamefont {Levis}, \citenamefont {Ybert},
  \citenamefont {Bocquet}, \citenamefont {Berthier},\ and\ \citenamefont
  {Cottin-Bizonne}}]{Ginot2015_PRX}%
  \BibitemOpen
  \bibfield  {author} {\bibinfo {author} {\bibfnamefont {F.}~\bibnamefont
  {Ginot}}, \bibinfo {author} {\bibfnamefont {I.}~\bibnamefont {Theurkauff}},
  \bibinfo {author} {\bibfnamefont {D.}~\bibnamefont {Levis}}, \bibinfo
  {author} {\bibfnamefont {C.}~\bibnamefont {Ybert}}, \bibinfo {author}
  {\bibfnamefont {L.}~\bibnamefont {Bocquet}}, \bibinfo {author} {\bibfnamefont
  {L.}~\bibnamefont {Berthier}}, \ and\ \bibinfo {author} {\bibfnamefont
  {C.}~\bibnamefont {Cottin-Bizonne}},\ }\href
  {https://doi.org/10.1103/PhysRevX.5.011004} {\bibfield  {journal} {\bibinfo
  {journal} {Phys. Rev. X}\ }\textbf {\bibinfo {volume} {5}},\ \bibinfo {pages}
  {011004} (\bibinfo {year} {2015})}\BibitemShut {NoStop}%
\bibitem [{\citenamefont {Speck}\ and\ \citenamefont
  {Jack}(2016)}]{Speck_PRE2016}%
  \BibitemOpen
  \bibfield  {author} {\bibinfo {author} {\bibfnamefont {T.}~\bibnamefont
  {Speck}}\ and\ \bibinfo {author} {\bibfnamefont {R.~L.}\ \bibnamefont
  {Jack}},\ }\href {https://doi.org/10.1103/PhysRevE.93.062605} {\bibfield
  {journal} {\bibinfo  {journal} {Phys. Rev. E}\ }\textbf {\bibinfo {volume}
  {93}},\ \bibinfo {pages} {062605} (\bibinfo {year} {2016})}\BibitemShut
  {NoStop}%
\bibitem [{\citenamefont {Marconi}, \citenamefont {Maggi},\ and\ \citenamefont
  {Melchionna}(2016)}]{Marconi2016_Pressure}%
  \BibitemOpen
  \bibfield  {author} {\bibinfo {author} {\bibfnamefont {U.~M.~B.}\
  \bibnamefont {Marconi}}, \bibinfo {author} {\bibfnamefont {C.}~\bibnamefont
  {Maggi}}, \ and\ \bibinfo {author} {\bibfnamefont {S.}~\bibnamefont
  {Melchionna}},\ }\href {https://doi.org/10.1039/C6SM00667A} {\bibfield
  {journal} {\bibinfo  {journal} {Soft Matter}\ }\textbf {\bibinfo {volume}
  {12}},\ \bibinfo {pages} {5727} (\bibinfo {year} {2016})}\BibitemShut
  {NoStop}%
\bibitem [{\citenamefont {Junot}\ \emph {et~al.}(2017)\citenamefont {Junot},
  \citenamefont {Briand}, \citenamefont {Ledesma-Alonso},\ and\ \citenamefont
  {Dauchot}}]{Junot_PRL2017}%
  \BibitemOpen
  \bibfield  {author} {\bibinfo {author} {\bibfnamefont {G.}~\bibnamefont
  {Junot}}, \bibinfo {author} {\bibfnamefont {G.}~\bibnamefont {Briand}},
  \bibinfo {author} {\bibfnamefont {R.}~\bibnamefont {Ledesma-Alonso}}, \ and\
  \bibinfo {author} {\bibfnamefont {O.}~\bibnamefont {Dauchot}},\ }\href
  {https://doi.org/10.1103/PhysRevLett.119.028002} {\bibfield  {journal}
  {\bibinfo  {journal} {Phys. Rev. Lett.}\ }\textbf {\bibinfo {volume} {119}},\
  \bibinfo {pages} {028002} (\bibinfo {year} {2017})}\BibitemShut {NoStop}%
\bibitem [{\citenamefont {Lee}(2017)}]{Lee_SM2017}%
  \BibitemOpen
  \bibfield  {author} {\bibinfo {author} {\bibfnamefont {C.~F.}\ \bibnamefont
  {Lee}},\ }\href {https://doi.org/10.1039/C6SM01978A} {\bibfield  {journal}
  {\bibinfo  {journal} {Soft Matter}\ }\textbf {\bibinfo {volume} {13}},\
  \bibinfo {pages} {376} (\bibinfo {year} {2017})}\BibitemShut {NoStop}%
\bibitem [{\citenamefont {Paoluzzi}\ \emph {et~al.}(2016)\citenamefont
  {Paoluzzi}, \citenamefont {{Di Leonardo}}, \citenamefont {Marchetti},\ and\
  \citenamefont {Angelani}}]{Paoluzzi_SR2016}%
  \BibitemOpen
  \bibfield  {author} {\bibinfo {author} {\bibfnamefont {M.}~\bibnamefont
  {Paoluzzi}}, \bibinfo {author} {\bibfnamefont {R.}~\bibnamefont {{Di
  Leonardo}}}, \bibinfo {author} {\bibfnamefont {M.~C.}\ \bibnamefont
  {Marchetti}}, \ and\ \bibinfo {author} {\bibfnamefont {L.}~\bibnamefont
  {Angelani}},\ }\href {https://doi.org/10.1038/srep34146} {\bibfield
  {journal} {\bibinfo  {journal} {Sci. Rep.}\ }\textbf {\bibinfo {volume}
  {6}},\ \bibinfo {pages} {34146} (\bibinfo {year} {2016})}\BibitemShut
  {NoStop}%
\bibitem [{\citenamefont {Fily}\ \emph {et~al.}(2018)\citenamefont {Fily},
  \citenamefont {Kafri}, \citenamefont {Solon}, \citenamefont {Tailleur},\ and\
  \citenamefont {Turner}}]{Fily_JPA2018}%
  \BibitemOpen
  \bibfield  {author} {\bibinfo {author} {\bibfnamefont {Y.}~\bibnamefont
  {Fily}}, \bibinfo {author} {\bibfnamefont {Y.}~\bibnamefont {Kafri}},
  \bibinfo {author} {\bibfnamefont {A.~P.}\ \bibnamefont {Solon}}, \bibinfo
  {author} {\bibfnamefont {J.}~\bibnamefont {Tailleur}}, \ and\ \bibinfo
  {author} {\bibfnamefont {A.}~\bibnamefont {Turner}},\ }\href
  {https://doi.org/10.1088/1751-8121/aa99b6} {\bibfield  {journal} {\bibinfo
  {journal} {J. Phys. A}\ }\textbf {\bibinfo {volume} {51}},\ \bibinfo {pages}
  {044003} (\bibinfo {year} {2018})}\BibitemShut {NoStop}%
\bibitem [{\citenamefont {Cates}\ \emph {et~al.}(2010)\citenamefont {Cates},
  \citenamefont {Marenduzzo}, \citenamefont {Pagonabarraga},\ and\
  \citenamefont {Tailleur}}]{Cates_PNAS2010}%
  \BibitemOpen
  \bibfield  {author} {\bibinfo {author} {\bibfnamefont {M.~E.}\ \bibnamefont
  {Cates}}, \bibinfo {author} {\bibfnamefont {D.}~\bibnamefont {Marenduzzo}},
  \bibinfo {author} {\bibfnamefont {I.}~\bibnamefont {Pagonabarraga}}, \ and\
  \bibinfo {author} {\bibfnamefont {J.}~\bibnamefont {Tailleur}},\ }\href
  {https://doi.org/10.1073/pnas.1001994107} {\bibfield  {journal} {\bibinfo
  {journal} {Proc. Natl. Acad. Sci. U.S.A.}\ }\textbf {\bibinfo {volume}
  {107}},\ \bibinfo {pages} {11715} (\bibinfo {year} {2010})}\BibitemShut
  {NoStop}%
\bibitem [{\citenamefont {Tjhung}, \citenamefont {Marenduzzo},\ and\
  \citenamefont {Cates}(2012)}]{Tjhung_PNAS2012}%
  \BibitemOpen
  \bibfield  {author} {\bibinfo {author} {\bibfnamefont {E.}~\bibnamefont
  {Tjhung}}, \bibinfo {author} {\bibfnamefont {D.}~\bibnamefont {Marenduzzo}},
  \ and\ \bibinfo {author} {\bibfnamefont {M.~E.}\ \bibnamefont {Cates}},\
  }\href {https://doi.org/10.1073/pnas.1200843109} {\bibfield  {journal}
  {\bibinfo  {journal} {Proc. Natl. Acad. Sci. U.S.A.}\ }\textbf {\bibinfo
  {volume} {109}},\ \bibinfo {pages} {12381} (\bibinfo {year}
  {2012})}\BibitemShut {NoStop}%
\bibitem [{\citenamefont {Tjhung}, \citenamefont {Cates},\ and\ \citenamefont
  {Marenduzzo}(2017)}]{Tjhung_PNAS2016}%
  \BibitemOpen
  \bibfield  {author} {\bibinfo {author} {\bibfnamefont {E.}~\bibnamefont
  {Tjhung}}, \bibinfo {author} {\bibfnamefont {M.~E.}\ \bibnamefont {Cates}}, \
  and\ \bibinfo {author} {\bibfnamefont {D.}~\bibnamefont {Marenduzzo}},\
  }\href {https://doi.org/10.1073/pnas.1619960114} {\bibfield  {journal}
  {\bibinfo  {journal} {Proc. Natl. Acad. Sci. U.S.A.}\ }\textbf {\bibinfo
  {volume} {114}},\ \bibinfo {pages} {4631} (\bibinfo {year}
  {2017})}\BibitemShut {NoStop}%
\bibitem [{\citenamefont {Vladescu}\ \emph {et~al.}(2014)\citenamefont
  {Vladescu}, \citenamefont {Marsden}, \citenamefont {Schwarz-Linek},
  \citenamefont {Martinez}, \citenamefont {Arlt}, \citenamefont {Morozov},
  \citenamefont {Marenduzzo}, \citenamefont {Cates},\ and\ \citenamefont
  {Poon}}]{Vladescu_PRL2014}%
  \BibitemOpen
  \bibfield  {author} {\bibinfo {author} {\bibfnamefont {I.~D.}\ \bibnamefont
  {Vladescu}}, \bibinfo {author} {\bibfnamefont {E.~J.}\ \bibnamefont
  {Marsden}}, \bibinfo {author} {\bibfnamefont {J.}~\bibnamefont
  {Schwarz-Linek}}, \bibinfo {author} {\bibfnamefont {V.~A.}\ \bibnamefont
  {Martinez}}, \bibinfo {author} {\bibfnamefont {J.}~\bibnamefont {Arlt}},
  \bibinfo {author} {\bibfnamefont {A.~N.}\ \bibnamefont {Morozov}}, \bibinfo
  {author} {\bibfnamefont {D.}~\bibnamefont {Marenduzzo}}, \bibinfo {author}
  {\bibfnamefont {M.~E.}\ \bibnamefont {Cates}}, \ and\ \bibinfo {author}
  {\bibfnamefont {W.~C.~K.}\ \bibnamefont {Poon}},\ }\href
  {https://doi.org/10.1103/PhysRevLett.113.268101} {\bibfield  {journal}
  {\bibinfo  {journal} {Phys. Rev. Lett.}\ }\textbf {\bibinfo {volume} {113}},\
  \bibinfo {pages} {268101} (\bibinfo {year} {2014})}\BibitemShut {NoStop}%
\bibitem [{\citenamefont {Hennes}\ \emph {et~al.}(2017)\citenamefont {Hennes},
  \citenamefont {Tailleur}, \citenamefont {Charron},\ and\ \citenamefont
  {Daerr}}]{Hennes_PNAS2017}%
  \BibitemOpen
  \bibfield  {author} {\bibinfo {author} {\bibfnamefont {M.}~\bibnamefont
  {Hennes}}, \bibinfo {author} {\bibfnamefont {J.}~\bibnamefont {Tailleur}},
  \bibinfo {author} {\bibfnamefont {G.}~\bibnamefont {Charron}}, \ and\
  \bibinfo {author} {\bibfnamefont {A.}~\bibnamefont {Daerr}},\ }\href
  {https://doi.org/10.1073/pnas.1703997114} {\bibfield  {journal} {\bibinfo
  {journal} {Proc. Natl. Acad. Sci. U.S.A.}\ }\textbf {\bibinfo {volume}
  {114}},\ \bibinfo {pages} {5958} (\bibinfo {year} {2017})}\BibitemShut
  {NoStop}%
\bibitem [{\citenamefont {Jin}\ \emph {et~al.}(2018)\citenamefont {Jin},
  \citenamefont {Hokmabad}, \citenamefont {Baldwin},\ and\ \citenamefont
  {Maass}}]{Jin_2018}%
  \BibitemOpen
  \bibfield  {author} {\bibinfo {author} {\bibfnamefont {C.}~\bibnamefont
  {Jin}}, \bibinfo {author} {\bibfnamefont {B.~V.}\ \bibnamefont {Hokmabad}},
  \bibinfo {author} {\bibfnamefont {K.~A.}\ \bibnamefont {Baldwin}}, \ and\
  \bibinfo {author} {\bibfnamefont {C.~C.}\ \bibnamefont {Maass}},\ }\href
  {https://doi.org/10.1088/1361-648X/aaa208} {\bibfield  {journal} {\bibinfo
  {journal} {J. Phys.: Condens. Matter}\ }\textbf {\bibinfo {volume} {30}},\
  \bibinfo {pages} {054003} (\bibinfo {year} {2018})}\BibitemShut {NoStop}%
\bibitem [{\citenamefont {Zwicker}\ \emph {et~al.}(2016)\citenamefont
  {Zwicker}, \citenamefont {Seyboldt}, \citenamefont {Weber}, \citenamefont
  {Hyman},\ and\ \citenamefont {J{\"u}licher}}]{Julicher:Nature2016}%
  \BibitemOpen
  \bibfield  {author} {\bibinfo {author} {\bibfnamefont {D.}~\bibnamefont
  {Zwicker}}, \bibinfo {author} {\bibfnamefont {R.}~\bibnamefont {Seyboldt}},
  \bibinfo {author} {\bibfnamefont {C.~A.}\ \bibnamefont {Weber}}, \bibinfo
  {author} {\bibfnamefont {A.~A.}\ \bibnamefont {Hyman}}, \ and\ \bibinfo
  {author} {\bibfnamefont {F.}~\bibnamefont {J{\"u}licher}},\ }\href
  {https://doi.org/10.1038/nphys3984} {\bibfield  {journal} {\bibinfo
  {journal} {Nat. Phys.}\ }\textbf {\bibinfo {volume} {13}},\ \bibinfo {pages}
  {408} (\bibinfo {year} {2016})}\BibitemShut {NoStop}%
\bibitem [{\citenamefont {Zwicker}, \citenamefont {Hyman},\ and\ \citenamefont
  {J{\"u}licher}(2015)}]{Julicher_PRE2015}%
  \BibitemOpen
  \bibfield  {author} {\bibinfo {author} {\bibfnamefont {D.}~\bibnamefont
  {Zwicker}}, \bibinfo {author} {\bibfnamefont {A.~A.}\ \bibnamefont {Hyman}},
  \ and\ \bibinfo {author} {\bibfnamefont {F.}~\bibnamefont {J{\"u}licher}},\
  }\href {https://doi.org/10.1103/PhysRevE.92.012317} {\bibfield  {journal}
  {\bibinfo  {journal} {Phys. Rev. E}\ }\textbf {\bibinfo {volume} {92}},\
  \bibinfo {pages} {012317} (\bibinfo {year} {2015})}\BibitemShut {NoStop}%
\bibitem [{\citenamefont {Weber}, \citenamefont {Lee},\ and\ \citenamefont
  {J{\"u}licher}(2017)}]{Julicher_NJP2017}%
  \BibitemOpen
  \bibfield  {author} {\bibinfo {author} {\bibfnamefont {C.~A.}\ \bibnamefont
  {Weber}}, \bibinfo {author} {\bibfnamefont {C.~F.}\ \bibnamefont {Lee}}, \
  and\ \bibinfo {author} {\bibfnamefont {F.}~\bibnamefont {J{\"u}licher}},\
  }\href {https://doi.org/10.1088/1367-2630/aa6b84} {\bibfield  {journal}
  {\bibinfo  {journal} {New J. Phys.}\ }\textbf {\bibinfo {volume} {19}},\
  \bibinfo {pages} {053021} (\bibinfo {year} {2017})}\BibitemShut {NoStop}%
\bibitem [{\citenamefont {Golestanian}(2016)}]{Golestanian_Nature2017}%
  \BibitemOpen
  \bibfield  {author} {\bibinfo {author} {\bibfnamefont {R.}~\bibnamefont
  {Golestanian}},\ }\href {https://doi.org/10.1038/nphys3998} {\bibfield
  {journal} {\bibinfo  {journal} {Nat. Phys.}\ }\textbf {\bibinfo {volume}
  {13}},\ \bibinfo {pages} {323} (\bibinfo {year} {2016})}\BibitemShut
  {NoStop}%
\bibitem [{\citenamefont {Cates}\ and\ \citenamefont
  {Tjhung}(2018)}]{Cate_2018review}%
  \BibitemOpen
  \bibfield  {author} {\bibinfo {author} {\bibfnamefont {M.~E.}\ \bibnamefont
  {Cates}}\ and\ \bibinfo {author} {\bibfnamefont {E.}~\bibnamefont {Tjhung}},\
  }\href {https://doi.org/10.1017/jfm.2017.832} {\bibfield  {journal} {\bibinfo
   {journal} {J. Fluid Mech.}\ }\textbf {\bibinfo {volume} {836}},\ \bibinfo
  {pages} {P1} (\bibinfo {year} {2018})}\BibitemShut {NoStop}%
\bibitem [{\citenamefont {Israelachvili}(2011)}]{Israelachvili2015}%
  \BibitemOpen
  \bibfield  {author} {\bibinfo {author} {\bibfnamefont {J.~N.}\ \bibnamefont
  {Israelachvili}},\ }\href
  {https://books.google.com/books/about/Intermolecular_and_Surface_Forces.html?id=MVbWBhubrgIC}
  {\emph {\bibinfo {title} {{Intermolecular and Surface Forces}}}},\ \bibinfo
  {edition} {3rd}\ ed.\ (\bibinfo  {publisher} {Academic Press},\ \bibinfo
  {address} {Amsterdam},\ \bibinfo {year} {2011})\BibitemShut {NoStop}%
\bibitem [{\citenamefont {de~Gennes}, \citenamefont {Brochard-Wyart},\ and\
  \citenamefont {Quere}(2004)}]{deGennes_Capillarity_2004}%
  \BibitemOpen
  \bibfield  {author} {\bibinfo {author} {\bibfnamefont {P.-G.}\ \bibnamefont
  {de~Gennes}}, \bibinfo {author} {\bibfnamefont {F.}~\bibnamefont
  {Brochard-Wyart}}, \ and\ \bibinfo {author} {\bibfnamefont {D.}~\bibnamefont
  {Quere}},\ }\href {https://doi.org/10.1007/978-0-387-21656-0} {\emph
  {\bibinfo {title} {{Capillarity and Wetting Phenomena}}}}\ (\bibinfo
  {publisher} {Springer},\ \bibinfo {address} {New York},\ \bibinfo {year}
  {2004})\BibitemShut {NoStop}%
\bibitem [{\citenamefont {Ostwald}(1900)}]{Ostwald1900}%
  \BibitemOpen
  \bibfield  {author} {\bibinfo {author} {\bibfnamefont {W.}~\bibnamefont
  {Ostwald}},\ }\href {https://doi.org/10.1515/zpch-1900-3431} {\bibfield
  {journal} {\bibinfo  {journal} {Z. Phys. Chem.}\ }\textbf {\bibinfo {volume}
  {34}},\ \bibinfo {pages} {495} (\bibinfo {year} {1900})}\BibitemShut
  {NoStop}%
\bibitem [{\citenamefont {{Hyman A. A.}}, \citenamefont {{Weber C. A.}},\ and\
  \citenamefont {{J{\"u}licher F.}}(2014)}]{Hyman_2014}%
  \BibitemOpen
  \bibfield  {author} {\bibinfo {author} {\bibnamefont {{Hyman A. A.}}},
  \bibinfo {author} {\bibnamefont {{Weber C. A.}}}, \ and\ \bibinfo {author}
  {\bibnamefont {{J{\"u}licher F.}}},\ }\href
  {https://doi.org/10.1146/annurev-cellbio-100913-013325} {\bibfield  {journal}
  {\bibinfo  {journal} {Annu. Rev. Cell Dev. Biol.}\ }\textbf {\bibinfo
  {volume} {30}},\ \bibinfo {pages} {39} (\bibinfo {year} {2014})}\BibitemShut
  {NoStop}%
\bibitem [{\citenamefont {Gonnella}\ \emph {et~al.}(2015)\citenamefont
  {Gonnella}, \citenamefont {Marenduzzo}, \citenamefont {Suma},\ and\
  \citenamefont {Tiribocchi}}]{Marenduzzo_MIPS2015}%
  \BibitemOpen
  \bibfield  {author} {\bibinfo {author} {\bibfnamefont {G.}~\bibnamefont
  {Gonnella}}, \bibinfo {author} {\bibfnamefont {D.}~\bibnamefont
  {Marenduzzo}}, \bibinfo {author} {\bibfnamefont {A.}~\bibnamefont {Suma}}, \
  and\ \bibinfo {author} {\bibfnamefont {A.}~\bibnamefont {Tiribocchi}},\
  }\href {https://doi.org/10.1016/j.crhy.2015.05.001} {\bibfield  {journal}
  {\bibinfo  {journal} {C. R. Phys.}\ }\textbf {\bibinfo {volume} {16}},\
  \bibinfo {pages} {316} (\bibinfo {year} {2015})}\BibitemShut {NoStop}%
\bibitem [{\citenamefont {Wittkowski}\ \emph {et~al.}(2014)\citenamefont
  {Wittkowski}, \citenamefont {Tiribocchi}, \citenamefont {Stenhammar},
  \citenamefont {Allen}, \citenamefont {Marenduzzo},\ and\ \citenamefont
  {Cates}}]{Cates_Nature2014}%
  \BibitemOpen
  \bibfield  {author} {\bibinfo {author} {\bibfnamefont {R.}~\bibnamefont
  {Wittkowski}}, \bibinfo {author} {\bibfnamefont {A.}~\bibnamefont
  {Tiribocchi}}, \bibinfo {author} {\bibfnamefont {J.}~\bibnamefont
  {Stenhammar}}, \bibinfo {author} {\bibfnamefont {R.~J.}\ \bibnamefont
  {Allen}}, \bibinfo {author} {\bibfnamefont {D.}~\bibnamefont {Marenduzzo}}, \
  and\ \bibinfo {author} {\bibfnamefont {M.~E.}\ \bibnamefont {Cates}},\ }\href
  {https://doi.org/10.1038/ncomms5351} {\bibfield  {journal} {\bibinfo
  {journal} {Nat. Commun.}\ }\textbf {\bibinfo {volume} {5}},\ \bibinfo {pages}
  {4351} (\bibinfo {year} {2014})}\BibitemShut {NoStop}%
\bibitem [{\citenamefont {Tailleur}\ and\ \citenamefont
  {Cates}(2008)}]{Tailleur2008_StatisticalMechanics}%
  \BibitemOpen
  \bibfield  {author} {\bibinfo {author} {\bibfnamefont {J.}~\bibnamefont
  {Tailleur}}\ and\ \bibinfo {author} {\bibfnamefont {M.~E.}\ \bibnamefont
  {Cates}},\ }\href {https://doi.org/10.1103/PhysRevLett.100.218103} {\bibfield
   {journal} {\bibinfo  {journal} {Phys. Rev. Lett.}\ }\textbf {\bibinfo
  {volume} {100}},\ \bibinfo {pages} {218103} (\bibinfo {year}
  {2008})}\BibitemShut {NoStop}%
\bibitem [{\citenamefont {Fily}\ and\ \citenamefont
  {Marchetti}(2012)}]{Fily2012_PhaseSeparation}%
  \BibitemOpen
  \bibfield  {author} {\bibinfo {author} {\bibfnamefont {Y.}~\bibnamefont
  {Fily}}\ and\ \bibinfo {author} {\bibfnamefont {M.~C.}\ \bibnamefont
  {Marchetti}},\ }\href {https://doi.org/10.1103/PhysRevLett.108.235702}
  {\bibfield  {journal} {\bibinfo  {journal} {Phys. Rev. Lett.}\ }\textbf
  {\bibinfo {volume} {108}},\ \bibinfo {pages} {235702} (\bibinfo {year}
  {2012})}\BibitemShut {NoStop}%
\bibitem [{\citenamefont {Redner}, \citenamefont {Hagan},\ and\ \citenamefont
  {Baskaran}(2013)}]{Redner2013_Structure}%
  \BibitemOpen
  \bibfield  {author} {\bibinfo {author} {\bibfnamefont {G.~S.}\ \bibnamefont
  {Redner}}, \bibinfo {author} {\bibfnamefont {M.~F.}\ \bibnamefont {Hagan}}, \
  and\ \bibinfo {author} {\bibfnamefont {A.}~\bibnamefont {Baskaran}},\ }\href
  {https://doi.org/10.1103/PhysRevLett.110.055701} {\bibfield  {journal}
  {\bibinfo  {journal} {Phys. Rev. Lett.}\ }\textbf {\bibinfo {volume} {110}},\
  \bibinfo {pages} {055701} (\bibinfo {year} {2013})}\BibitemShut {NoStop}%
\bibitem [{\citenamefont {Buttinoni}\ \emph {et~al.}(2013)\citenamefont
  {Buttinoni}, \citenamefont {Bialk{\'e}}, \citenamefont {K{\"u}mmel},
  \citenamefont {L{\"o}wen}, \citenamefont {Bechinger},\ and\ \citenamefont
  {Speck}}]{Buttinoni2013_DynamicalClustering}%
  \BibitemOpen
  \bibfield  {author} {\bibinfo {author} {\bibfnamefont {I.}~\bibnamefont
  {Buttinoni}}, \bibinfo {author} {\bibfnamefont {J.}~\bibnamefont
  {Bialk{\'e}}}, \bibinfo {author} {\bibfnamefont {F.}~\bibnamefont
  {K{\"u}mmel}}, \bibinfo {author} {\bibfnamefont {H.}~\bibnamefont
  {L{\"o}wen}}, \bibinfo {author} {\bibfnamefont {C.}~\bibnamefont
  {Bechinger}}, \ and\ \bibinfo {author} {\bibfnamefont {T.}~\bibnamefont
  {Speck}},\ }\href {https://doi.org/10.1103/PhysRevLett.110.238301} {\bibfield
   {journal} {\bibinfo  {journal} {Phys. Rev. Lett.}\ }\textbf {\bibinfo
  {volume} {110}},\ \bibinfo {pages} {238301} (\bibinfo {year}
  {2013})}\BibitemShut {NoStop}%
\bibitem [{\citenamefont {Cates}\ and\ \citenamefont
  {Tailleur}(2015)}]{Cates2015_Motility-Induced}%
  \BibitemOpen
  \bibfield  {author} {\bibinfo {author} {\bibfnamefont {M.~E.}\ \bibnamefont
  {Cates}}\ and\ \bibinfo {author} {\bibfnamefont {J.}~\bibnamefont
  {Tailleur}},\ }\href
  {https://doi.org/10.1146/annurev-conmatphys-031214-014710} {\bibfield
  {journal} {\bibinfo  {journal} {Annu. Rev. Condens. Matter Phys.}\ }\textbf
  {\bibinfo {volume} {6}},\ \bibinfo {pages} {219} (\bibinfo {year}
  {2015})}\BibitemShut {NoStop}%
\bibitem [{\citenamefont {Liu}\ \emph {et~al.}(2016)\citenamefont {Liu},
  \citenamefont {Rietkerk}, \citenamefont {Herman}, \citenamefont {Piersma},
  \citenamefont {Fryxell},\ and\ \citenamefont {van~de Koppel}}]{Liu2016}%
  \BibitemOpen
  \bibfield  {author} {\bibinfo {author} {\bibfnamefont {Q.-X.}\ \bibnamefont
  {Liu}}, \bibinfo {author} {\bibfnamefont {M.}~\bibnamefont {Rietkerk}},
  \bibinfo {author} {\bibfnamefont {P.~M.}\ \bibnamefont {Herman}}, \bibinfo
  {author} {\bibfnamefont {T.}~\bibnamefont {Piersma}}, \bibinfo {author}
  {\bibfnamefont {J.~M.}\ \bibnamefont {Fryxell}}, \ and\ \bibinfo {author}
  {\bibfnamefont {J.}~\bibnamefont {van~de Koppel}},\ }\href
  {https://doi.org/10.1016/j.plrev.2016.07.009} {\bibfield  {journal} {\bibinfo
   {journal} {Phys. Life Rev.}\ }\textbf {\bibinfo {volume} {19}},\ \bibinfo
  {pages} {107} (\bibinfo {year} {2016})}\BibitemShut {NoStop}%
\bibitem [{\citenamefont {Stenhammar}\ \emph {et~al.}(2015)\citenamefont
  {Stenhammar}, \citenamefont {Wittkowski}, \citenamefont {Marenduzzo},\ and\
  \citenamefont {Cates}}]{Stenhammar_PRL2015}%
  \BibitemOpen
  \bibfield  {author} {\bibinfo {author} {\bibfnamefont {J.}~\bibnamefont
  {Stenhammar}}, \bibinfo {author} {\bibfnamefont {R.}~\bibnamefont
  {Wittkowski}}, \bibinfo {author} {\bibfnamefont {D.}~\bibnamefont
  {Marenduzzo}}, \ and\ \bibinfo {author} {\bibfnamefont {M.~E.}\ \bibnamefont
  {Cates}},\ }\href {https://doi.org/10.1103/PhysRevLett.114.018301} {\bibfield
   {journal} {\bibinfo  {journal} {Phys. Rev. Lett.}\ }\textbf {\bibinfo
  {volume} {114}},\ \bibinfo {pages} {018301} (\bibinfo {year}
  {2015})}\BibitemShut {NoStop}%
\bibitem [{\citenamefont {Wittkowski}, \citenamefont {Stenhammar},\ and\
  \citenamefont {Cates}(2017)}]{Wittkowski_NJP2017}%
  \BibitemOpen
  \bibfield  {author} {\bibinfo {author} {\bibfnamefont {R.}~\bibnamefont
  {Wittkowski}}, \bibinfo {author} {\bibfnamefont {J.}~\bibnamefont
  {Stenhammar}}, \ and\ \bibinfo {author} {\bibfnamefont {M.~E.}\ \bibnamefont
  {Cates}},\ }\href {https://doi.org/10.1088/1367-2630/aa8195} {\bibfield
  {journal} {\bibinfo  {journal} {New J. Phys.}\ }\textbf {\bibinfo {volume}
  {19}},\ \bibinfo {pages} {105003} (\bibinfo {year} {2017})}\BibitemShut
  {NoStop}%
\bibitem [{\citenamefont {Ebbens}\ \emph {et~al.}(2010)\citenamefont {Ebbens},
  \citenamefont {Jones}, \citenamefont {Ryan}, \citenamefont {Golestanian},\
  and\ \citenamefont {Howse}}]{Golestanian:PRE2010}%
  \BibitemOpen
  \bibfield  {author} {\bibinfo {author} {\bibfnamefont {S.}~\bibnamefont
  {Ebbens}}, \bibinfo {author} {\bibfnamefont {R.~A.~L.}\ \bibnamefont
  {Jones}}, \bibinfo {author} {\bibfnamefont {A.~J.}\ \bibnamefont {Ryan}},
  \bibinfo {author} {\bibfnamefont {R.}~\bibnamefont {Golestanian}}, \ and\
  \bibinfo {author} {\bibfnamefont {J.~R.}\ \bibnamefont {Howse}},\ }\href
  {https://doi.org/10.1103/PhysRevE.82.015304} {\bibfield  {journal} {\bibinfo
  {journal} {Phys. Rev. E}\ }\textbf {\bibinfo {volume} {82}},\ \bibinfo
  {pages} {015304} (\bibinfo {year} {2010})}\BibitemShut {NoStop}%
\bibitem [{\citenamefont {Takagi}\ \emph {et~al.}(2014)\citenamefont {Takagi},
  \citenamefont {Palacci}, \citenamefont {Braunschweig}, \citenamefont
  {Shelley},\ and\ \citenamefont {Zhang}}]{takagi}%
  \BibitemOpen
  \bibfield  {author} {\bibinfo {author} {\bibfnamefont {D.}~\bibnamefont
  {Takagi}}, \bibinfo {author} {\bibfnamefont {J.}~\bibnamefont {Palacci}},
  \bibinfo {author} {\bibfnamefont {A.~B.}\ \bibnamefont {Braunschweig}},
  \bibinfo {author} {\bibfnamefont {M.~J.}\ \bibnamefont {Shelley}}, \ and\
  \bibinfo {author} {\bibfnamefont {J.}~\bibnamefont {Zhang}},\ }\href
  {https://doi.org/10.1039/C3SM52815D} {\bibfield  {journal} {\bibinfo
  {journal} {Soft Matter}\ }\textbf {\bibinfo {volume} {10}},\ \bibinfo {pages}
  {1784} (\bibinfo {year} {2014})}\BibitemShut {NoStop}%
\bibitem [{\citenamefont {Takagi}\ \emph {et~al.}(2013)\citenamefont {Takagi},
  \citenamefont {Braunschweig}, \citenamefont {Zhang},\ and\ \citenamefont
  {Shelley}}]{takagi2013}%
  \BibitemOpen
  \bibfield  {author} {\bibinfo {author} {\bibfnamefont {D.}~\bibnamefont
  {Takagi}}, \bibinfo {author} {\bibfnamefont {A.~B.}\ \bibnamefont
  {Braunschweig}}, \bibinfo {author} {\bibfnamefont {J.}~\bibnamefont {Zhang}},
  \ and\ \bibinfo {author} {\bibfnamefont {M.~J.}\ \bibnamefont {Shelley}},\
  }\href {https://doi.org/10.1103/PhysRevLett.110.038301} {\bibfield  {journal}
  {\bibinfo  {journal} {Phys. Rev. Lett.}\ }\textbf {\bibinfo {volume} {110}},\
  \bibinfo {pages} {038301} (\bibinfo {year} {2013})}\BibitemShut {NoStop}%
\bibitem [{\citenamefont {K{\"u}mmel}\ \emph {et~al.}(2013)\citenamefont
  {K{\"u}mmel}, \citenamefont {ten Hagen}, \citenamefont {Wittkowski},
  \citenamefont {Buttinoni}, \citenamefont {Eichhorn}, \citenamefont {Volpe},
  \citenamefont {L{\"o}wen},\ and\ \citenamefont
  {Bechinger}}]{Bechinger:PRL2013}%
  \BibitemOpen
  \bibfield  {author} {\bibinfo {author} {\bibfnamefont {F.}~\bibnamefont
  {K{\"u}mmel}}, \bibinfo {author} {\bibfnamefont {B.}~\bibnamefont {ten
  Hagen}}, \bibinfo {author} {\bibfnamefont {R.}~\bibnamefont {Wittkowski}},
  \bibinfo {author} {\bibfnamefont {I.}~\bibnamefont {Buttinoni}}, \bibinfo
  {author} {\bibfnamefont {R.}~\bibnamefont {Eichhorn}}, \bibinfo {author}
  {\bibfnamefont {G.}~\bibnamefont {Volpe}}, \bibinfo {author} {\bibfnamefont
  {H.}~\bibnamefont {L{\"o}wen}}, \ and\ \bibinfo {author} {\bibfnamefont
  {C.}~\bibnamefont {Bechinger}},\ }\href
  {https://doi.org/10.1103/PhysRevLett.110.198302} {\bibfield  {journal}
  {\bibinfo  {journal} {Phys. Rev. Lett.}\ }\textbf {\bibinfo {volume} {110}},\
  \bibinfo {pages} {198302} (\bibinfo {year} {2013})}\BibitemShut {NoStop}%
\bibitem [{\citenamefont {Ao}\ \emph {et~al.}(2014)\citenamefont {Ao},
  \citenamefont {Ghosh}, \citenamefont {Li}, \citenamefont {Schmid},
  \citenamefont {H{\"a}nggi},\ and\ \citenamefont
  {Marchesoni}}]{Xue:EPJST2014}%
  \BibitemOpen
  \bibfield  {author} {\bibinfo {author} {\bibfnamefont {X.}~\bibnamefont
  {Ao}}, \bibinfo {author} {\bibfnamefont {P.~K.}\ \bibnamefont {Ghosh}},
  \bibinfo {author} {\bibfnamefont {Y.}~\bibnamefont {Li}}, \bibinfo {author}
  {\bibfnamefont {G.}~\bibnamefont {Schmid}}, \bibinfo {author} {\bibfnamefont
  {P.}~\bibnamefont {H{\"a}nggi}}, \ and\ \bibinfo {author} {\bibfnamefont
  {F.}~\bibnamefont {Marchesoni}},\ }\href
  {https://doi.org/10.1140/epjst/e2014-02329-1} {\bibfield  {journal} {\bibinfo
   {journal} {Eur. Phys. J. Spec. Top.}\ }\textbf {\bibinfo {volume} {223}},\
  \bibinfo {pages} {3227} (\bibinfo {year} {2014})}\BibitemShut {NoStop}%
\bibitem [{\citenamefont {Ao}\ \emph {et~al.}(2015)\citenamefont {Ao},
  \citenamefont {Ghosh}, \citenamefont {Li}, \citenamefont {Schmid},
  \citenamefont {H{\"a}nggi},\ and\ \citenamefont {Marchesoni}}]{Xue:EPL2015}%
  \BibitemOpen
  \bibfield  {author} {\bibinfo {author} {\bibfnamefont {X.}~\bibnamefont
  {Ao}}, \bibinfo {author} {\bibfnamefont {P.~K.}\ \bibnamefont {Ghosh}},
  \bibinfo {author} {\bibfnamefont {Y.}~\bibnamefont {Li}}, \bibinfo {author}
  {\bibfnamefont {G.}~\bibnamefont {Schmid}}, \bibinfo {author} {\bibfnamefont
  {P.}~\bibnamefont {H{\"a}nggi}}, \ and\ \bibinfo {author} {\bibfnamefont
  {F.}~\bibnamefont {Marchesoni}},\ }\href
  {https://doi.org/10.1209/0295-5075/109/10003} {\bibfield  {journal} {\bibinfo
   {journal} {EPL (Europhysics Letters)}\ }\textbf {\bibinfo {volume} {109}},\
  \bibinfo {pages} {10003} (\bibinfo {year} {2015})}\BibitemShut {NoStop}%
\bibitem [{\citenamefont {{Davies Wykes}}\ \emph {et~al.}(2016)\citenamefont
  {{Davies Wykes}}, \citenamefont {Palacci}, \citenamefont {Adachi},
  \citenamefont {Ristroph}, \citenamefont {Zhong}, \citenamefont {Ward},
  \citenamefont {Zhang},\ and\ \citenamefont {Shelley}}]{Wykes_2016}%
  \BibitemOpen
  \bibfield  {author} {\bibinfo {author} {\bibfnamefont {M.~S.}\ \bibnamefont
  {{Davies Wykes}}}, \bibinfo {author} {\bibfnamefont {J.}~\bibnamefont
  {Palacci}}, \bibinfo {author} {\bibfnamefont {T.}~\bibnamefont {Adachi}},
  \bibinfo {author} {\bibfnamefont {L.}~\bibnamefont {Ristroph}}, \bibinfo
  {author} {\bibfnamefont {X.}~\bibnamefont {Zhong}}, \bibinfo {author}
  {\bibfnamefont {M.~D.}\ \bibnamefont {Ward}}, \bibinfo {author}
  {\bibfnamefont {J.}~\bibnamefont {Zhang}}, \ and\ \bibinfo {author}
  {\bibfnamefont {M.~J.}\ \bibnamefont {Shelley}},\ }\href
  {https://doi.org/10.1039/C5SM03127C} {\bibfield  {journal} {\bibinfo
  {journal} {Soft Matter}\ }\textbf {\bibinfo {volume} {12}},\ \bibinfo {pages}
  {4584} (\bibinfo {year} {2016})}\BibitemShut {NoStop}%
\bibitem [{\citenamefont {Nourhani}\ \emph {et~al.}(2013)\citenamefont
  {Nourhani}, \citenamefont {Lammert}, \citenamefont {Borhan},\ and\
  \citenamefont {Crespi}}]{Crespi:2013}%
  \BibitemOpen
  \bibfield  {author} {\bibinfo {author} {\bibfnamefont {A.}~\bibnamefont
  {Nourhani}}, \bibinfo {author} {\bibfnamefont {P.~E.}\ \bibnamefont
  {Lammert}}, \bibinfo {author} {\bibfnamefont {A.}~\bibnamefont {Borhan}}, \
  and\ \bibinfo {author} {\bibfnamefont {V.~H.}\ \bibnamefont {Crespi}},\
  }\href {https://doi.org/10.1103/PhysRevE.87.050301} {\bibfield  {journal}
  {\bibinfo  {journal} {Phys. Rev. E}\ }\textbf {\bibinfo {volume} {87}},\
  \bibinfo {pages} {050301(R)} (\bibinfo {year} {2013})}\BibitemShut {NoStop}%
\bibitem [{\citenamefont {van Teeffelen}\ and\ \citenamefont
  {L{\"o}wen}(2008)}]{Teeffelen:PRE2008}%
  \BibitemOpen
  \bibfield  {author} {\bibinfo {author} {\bibfnamefont {S.}~\bibnamefont {van
  Teeffelen}}\ and\ \bibinfo {author} {\bibfnamefont {H.}~\bibnamefont
  {L{\"o}wen}},\ }\href {https://doi.org/10.1103/PhysRevE.78.020101} {\bibfield
   {journal} {\bibinfo  {journal} {Phys. Rev. E}\ }\textbf {\bibinfo {volume}
  {78}},\ \bibinfo {pages} {020101} (\bibinfo {year} {2008})}\BibitemShut
  {NoStop}%
\bibitem [{\citenamefont {Reichhardt}\ and\ \citenamefont {{Olson
  Reichhardt}}(2013)}]{Reichhardt:2013}%
  \BibitemOpen
  \bibfield  {author} {\bibinfo {author} {\bibfnamefont {C.}~\bibnamefont
  {Reichhardt}}\ and\ \bibinfo {author} {\bibfnamefont {C.~J.}\ \bibnamefont
  {{Olson Reichhardt}}},\ }\href {https://doi.org/10.1103/PhysRevE.88.042306}
  {\bibfield  {journal} {\bibinfo  {journal} {Phys. Rev. E}\ }\textbf {\bibinfo
  {volume} {88}},\ \bibinfo {pages} {042306} (\bibinfo {year}
  {2013})}\BibitemShut {NoStop}%
\bibitem [{\citenamefont {Mijalkov}\ and\ \citenamefont
  {Volpe}(2013)}]{Mijalkov:2015}%
  \BibitemOpen
  \bibfield  {author} {\bibinfo {author} {\bibfnamefont {M.}~\bibnamefont
  {Mijalkov}}\ and\ \bibinfo {author} {\bibfnamefont {G.}~\bibnamefont
  {Volpe}},\ }\href {https://doi.org/10.1039/C3SM27923E} {\bibfield  {journal}
  {\bibinfo  {journal} {Soft Matter}\ }\textbf {\bibinfo {volume} {9}},\
  \bibinfo {pages} {6376} (\bibinfo {year} {2013})}\BibitemShut {NoStop}%
\bibitem [{\citenamefont {L{\"o}wen}(2016)}]{Lowen:EPJST2016}%
  \BibitemOpen
  \bibfield  {author} {\bibinfo {author} {\bibfnamefont {H.}~\bibnamefont
  {L{\"o}wen}},\ }\href {https://doi.org/10.1140/epjst/e2016-60054-6}
  {\bibfield  {journal} {\bibinfo  {journal} {Eur. Phys. J. Spec. Top.}\
  }\textbf {\bibinfo {volume} {225}},\ \bibinfo {pages} {2319} (\bibinfo {year}
  {2016})}\BibitemShut {NoStop}%
\bibitem [{\citenamefont {Uspal}\ \emph {et~al.}(2016)\citenamefont {Uspal},
  \citenamefont {Popescu}, \citenamefont {Dietrich},\ and\ \citenamefont
  {Tasinkevych}}]{Popescu_PRL2016}%
  \BibitemOpen
  \bibfield  {author} {\bibinfo {author} {\bibfnamefont {W.~E.}\ \bibnamefont
  {Uspal}}, \bibinfo {author} {\bibfnamefont {M.~N.}\ \bibnamefont {Popescu}},
  \bibinfo {author} {\bibfnamefont {S.}~\bibnamefont {Dietrich}}, \ and\
  \bibinfo {author} {\bibfnamefont {M.}~\bibnamefont {Tasinkevych}},\ }\href
  {https://doi.org/10.1103/PhysRevLett.117.048002} {\bibfield  {journal}
  {\bibinfo  {journal} {Phys. Rev. Lett.}\ }\textbf {\bibinfo {volume} {117}},\
  \bibinfo {pages} {048002} (\bibinfo {year} {2016})}\BibitemShut {NoStop}%
\bibitem [{\citenamefont {Liebchen}, \citenamefont {Cates},\ and\ \citenamefont
  {Marenduzzo}(2016)}]{Liebchen_2016}%
  \BibitemOpen
  \bibfield  {author} {\bibinfo {author} {\bibfnamefont {B.}~\bibnamefont
  {Liebchen}}, \bibinfo {author} {\bibfnamefont {M.~E.}\ \bibnamefont {Cates}},
  \ and\ \bibinfo {author} {\bibfnamefont {D.}~\bibnamefont {Marenduzzo}},\
  }\href {https://doi.org/10.1039/C6SM01162D} {\bibfield  {journal} {\bibinfo
  {journal} {Soft Matter}\ }\textbf {\bibinfo {volume} {12}},\ \bibinfo {pages}
  {7259} (\bibinfo {year} {2016})}\BibitemShut {NoStop}%
\bibitem [{\citenamefont {Simmchen}\ \emph {et~al.}(2016)\citenamefont
  {Simmchen}, \citenamefont {Katuri}, \citenamefont {Uspal}, \citenamefont
  {Popescu}, \citenamefont {Tasinkevych},\ and\ \citenamefont
  {S{\'a}nchez}}]{Simmchen_Nature2016}%
  \BibitemOpen
  \bibfield  {author} {\bibinfo {author} {\bibfnamefont {J.}~\bibnamefont
  {Simmchen}}, \bibinfo {author} {\bibfnamefont {J.}~\bibnamefont {Katuri}},
  \bibinfo {author} {\bibfnamefont {W.~E.}\ \bibnamefont {Uspal}}, \bibinfo
  {author} {\bibfnamefont {M.~N.}\ \bibnamefont {Popescu}}, \bibinfo {author}
  {\bibfnamefont {M.}~\bibnamefont {Tasinkevych}}, \ and\ \bibinfo {author}
  {\bibfnamefont {S.}~\bibnamefont {S{\'a}nchez}},\ }\href
  {https://doi.org/10.1038/ncomms10598} {\bibfield  {journal} {\bibinfo
  {journal} {Nat. Commun.}\ }\textbf {\bibinfo {volume} {7}},\ \bibinfo {pages}
  {10598} (\bibinfo {year} {2016})}\BibitemShut {NoStop}%
\bibitem [{\citenamefont {Ni}, \citenamefont {{Cohen Stuart}},\ and\
  \citenamefont {Bolhuis}(2015)}]{Ni2015_Tunable}%
  \BibitemOpen
  \bibfield  {author} {\bibinfo {author} {\bibfnamefont {R.}~\bibnamefont
  {Ni}}, \bibinfo {author} {\bibfnamefont {M.~A.}\ \bibnamefont {{Cohen
  Stuart}}}, \ and\ \bibinfo {author} {\bibfnamefont {P.~G.}\ \bibnamefont
  {Bolhuis}},\ }\href {https://doi.org/10.1103/PhysRevLett.114.018302}
  {\bibfield  {journal} {\bibinfo  {journal} {Phys. Rev. Lett.}\ }\textbf
  {\bibinfo {volume} {114}},\ \bibinfo {pages} {018302} (\bibinfo {year}
  {2015})}\BibitemShut {NoStop}%
\bibitem [{\citenamefont {{Zaeifi Yamchi}}\ and\ \citenamefont
  {Naji}(2017)}]{Mahdi2017_ChiralSwimmers}%
  \BibitemOpen
  \bibfield  {author} {\bibinfo {author} {\bibfnamefont {M.}~\bibnamefont
  {{Zaeifi Yamchi}}}\ and\ \bibinfo {author} {\bibfnamefont {A.}~\bibnamefont
  {Naji}},\ }\href {https://doi.org/10.1063/1.5001505} {\bibfield  {journal}
  {\bibinfo  {journal} {J. Chem. Phys.}\ }\textbf {\bibinfo {volume} {147}},\
  \bibinfo {pages} {194901} (\bibinfo {year} {2017})}\BibitemShut {NoStop}%
\bibitem [{\citenamefont {Saintillan}\ and\ \citenamefont
  {Shelley}(2015)}]{saintillan2015bookchapter}%
  \BibitemOpen
  \bibfield  {author} {\bibinfo {author} {\bibfnamefont {D.}~\bibnamefont
  {Saintillan}}\ and\ \bibinfo {author} {\bibfnamefont {M.~J.}\ \bibnamefont
  {Shelley}},\ }in\ \href {https://doi.org/10.1007/978-1-4939-2065-5} {\emph
  {\bibinfo {booktitle} {{Complex Fluids in Biological Systems}}}}\ (\bibinfo
  {publisher} {Springer},\ \bibinfo {address} {New York},\ \bibinfo {year}
  {2015})\ pp.\ \bibinfo {pages} {319--355}\BibitemShut {NoStop}%
\bibitem [{\citenamefont {Risken}(1996)}]{Risken_FP_1996}%
  \BibitemOpen
  \bibfield  {author} {\bibinfo {author} {\bibfnamefont {H.}~\bibnamefont
  {Risken}},\ }\href {https://doi.org/10.1007/978-3-642-61544-3} {\emph
  {\bibinfo {title} {{The Fokker-Planck Equation}}}}\ (\bibinfo  {publisher}
  {Springer},\ \bibinfo {address} {Berlin},\ \bibinfo {year}
  {1996})\BibitemShut {NoStop}%
\bibitem [{foo({\natexlab{a}})}]{footnote1}%
  \BibitemOpen
  \href@noop {} {}\bibinfo {note} {Our consideration of torque-free boundaries
  implies that the swim pressure should be a state function
  \cite{Solon2015_Nature}, which indeed agrees with our numerical inspections,
  showing no dependence on the strength of the repulsive interfacial potential,
  or its functional form (e.g., linear or harmonic).}\BibitemShut {Stop}%
\bibitem [{\citenamefont {Brennen}(2013)}]{Brennen_2013}%
  \BibitemOpen
  \bibfield  {author} {\bibinfo {author} {\bibfnamefont {C.~E.}\ \bibnamefont
  {Brennen}},\ }\href {https://doi.org/10.1017/CBO9781107338760} {\emph
  {\bibinfo {title} {{Cavitation and Bubble Dynamics}}}}\ (\bibinfo
  {publisher} {Cambridge University Press},\ \bibinfo {address} {Cambridge},\
  \bibinfo {year} {2013})\BibitemShut {NoStop}%
\bibitem [{foo({\natexlab{b}})}]{footnote2}%
  \BibitemOpen
  \href@noop {} {}\bibinfo {note} {Our numerical results for the higher-order
  moments up to $l=4$ confirm that the moments vanish everywhere except at the
  vicinity of the inclusion surface. Also, the higher moments take increasingly
  smaller values; e.g., $\textrm{max}|m_4|/\textrm{max}|m_2|\lesssim 0.25$ for
  $\tilde R=10$, $\Pe=10$.}\BibitemShut {Stop}%
\end{thebibliography}%

\end{document}